\documentclass[pdflatex,sn-mathphys-num]{sn-jnl}

\usepackage{graphicx}%
\usepackage{multirow}%
\usepackage{amsmath,amssymb,amsfonts}%
\usepackage{amsthm}%
\usepackage{mathrsfs}%
\usepackage[title]{appendix}%
\usepackage{xcolor}%
\usepackage{textcomp}%
\usepackage{manyfoot}%
\usepackage{booktabs}%
\usepackage{algorithm}%
\usepackage{algorithmicx}%
\usepackage{algpseudocode}%
\usepackage{listings}%
\usepackage{longtable}   
\usepackage{booktabs}    
\usepackage{amssymb}     
\usepackage{array}  

\theoremstyle{thmstyleone}%
\theoremstyle{thmstyletwo}%

\theoremstyle{thmstylethree}%

\begin{document}

\title[Article Title]{A Multi-Viewpoint Modeling Framework for Digital Twin Integration and Reuse with LLM-Assisted Compatibility Analysis}

\author[1]{\fnm{Nafiseh} \sur{Soveizi}}
\email{n.soveizi@uva.nl}

\author[1]{\fnm{Milan} \sur{Kopp}}
\email{milan.kopp@student.uva.nl}

\author[3]{\fnm{Parinaz} \sur{Rashidi}}
\email{p.rashidi@uva.nl}

\author[5]{\fnm{Qing} \sur{Zhan}}
\email{qing.zhan@nioz.nl}

\author[5]{\fnm{Geerten M.} \sur{Hengeveld}}
\email{G.Hengeveld@nioo.knaw.nl}

\author[6]{\fnm{Ioannis N.} \sur{Athanasiadis}}
\email{ioannis.athanasiadis@wur.nl}

\author*[1,2]{\fnm{Zhiming} \sur{Zhao}}
\email{z.zhao@uva.nl}

\affil*[1]{%
  \orgdiv{Faculty of Science},
  \orgname{University of Amsterdam},
  \orgaddress{\city{Amsterdam}, \country{The Netherlands}}
}

\affil[2]{%
  \orgdiv{LifeWatch ERIC Virtual Lab Innovation Center (VLIC)},
  \orgname{University of Amsterdam},
  \orgaddress{\city{Amsterdam}, \country{The Netherlands}}
}

\affil[3]{%
  \orgdiv{Information Technology Group},
  \orgname{University of Amsterdam},
  \orgaddress{\city{Amsterdam}, \country{The Netherlands}}
}

\affil[4]{%
  \orgdiv{Department of Terrestrial Ecology},
  \orgname{Netherlands Institute of Ecology (NIOO-KNAW)},
  \orgaddress{\city{Wageningen}, \country{The Netherlands}}
}

\affil[5]{%
  \orgdiv{Department of Coastal Systems},
  \orgname{Royal Netherlands Institute for Sea Research (NIOZ)},
  \orgaddress{\city{Texel}, \country{The Netherlands}}
}

\affil[6]{%
  \orgdiv{Institute for Biodiversity and Ecosystem Dynamics (IBED)},
  \orgname{Wageningen University},
  \orgaddress{\city{Wageningen}, \country{The Netherlands}}
}


\abstract{

Digital Twin (DT) ecosystems integrate heterogeneous computational models to represent complex systems under evolving, purpose-specific objectives. Systematic reuse of existing high-quality models and datasets is essential for scalable DT development, yet reuse is constrained by heterogeneity across semantic intent, data structures, behavioral interfaces, and execution environments. As a result, integration becomes a cross-model, cross-view consistency problem that is difficult to predict, quantify, and compare across alternative design choices. Existing standards and integration platforms address these concerns separately and provide limited support for structured, purpose-aware compatibility assessment and early feasibility analysis when models are reused under new DT objectives. This paper introduces a multi-viewpoint integration modeling framework grounded in the Reference Model of Open Distributed Processing (RM-ODP). The framework structures integration-relevant knowledge across domain, information, computational, engineering, and technology viewpoints, representing cross-view dependencies as explicit, machine-actionable metadata. It comprises (i) a viewpoint-structured Model Metamodel supporting systematic model description and discovery, and (ii) a pattern-aware Mismatch Detector that operationalizes cross-view compatibility constraints induced by integration patterns through a hybrid mechanism combining deterministic rule generation with Large Language Model (LLM)–assisted reasoning. This enables systematic identification of semantic, informational, and runtime inconsistencies and supports reasoning about integration feasibility and adaptation effort prior to implementation. Expert validation and an environmental modeling case study demonstrate that the approach enables structured compatibility reasoning, improves transparency of integration assumptions, strengthens cross-view interoperability, and supports systematic and scalable reuse in heterogeneous DT ecosystems\footnote{\url{https://github.com/nafisesoezy/DigitalTwinIntegration/}}.

}




\maketitle

\section{Introduction}
\label{Introduction}
Solving major scientific and engineering challenges—ranging from environmental sustainability to precision medicine—requires the integration of heterogeneous computational models, large datasets, and distributed infrastructures~\cite{vermeulen2020supporting,tuli2020next}. Digital Twins (DTs) have emerged as a paradigm for constructing continuously evolving digital representations of physical or conceptual systems~\cite{barricelli2019survey,tao2022digital}. From a software and systems modeling perspective, DTs constitute complex, multi-model systems in which independently developed artifacts must interoperate coherently to support a specific, human-defined purpose~\cite{bezivin2006model,blair2009models,bruel2010viewpoint}.

As DT objectives evolve, systematic reuse and repurposing of existing high-quality models and datasets become essential for scalable and sustainable development. However, reuse is fundamentally constrained by heterogeneity across semantic intent, information structures, behavioral interfaces, and execution environments. These differences arise from purpose-driven design decisions and stakeholder-specific modeling abstractions that shape assumptions, data representations, and interaction mechanisms~\cite{atkinson2003mda,france2007model}. Consequently, integration becomes a cross-model consistency problem spanning multiple abstraction layers, often addressed manually and in an ad hoc manner~\cite{zhao2022notebook,egyed2007fixing}.

Existing modeling and integration platforms—such as component-based frameworks, workflow systems, and Virtual Research Environments (VREs)—primarily focus on execution and data exchange. While they provide useful infrastructure, they offer limited support for systematically reasoning about (i) whether independently developed models are mutually compatible and (ii) whether they align with the objectives of a new DT. In practice, successful integration requires consistency across multiple levels, including domain semantics, data structures, behavioral interfaces, and runtime coordination, all aligned with the intended DT goals. However, such full alignment is rare in real-world reuse scenarios. Instead, integration involves resolving mismatches across these levels, requiring non-trivial engineering effort. This difficulty arises because integration decisions depend on interrelated constraints that span multiple viewpoints and are often only revealed late in implementation, making integration not only technically challenging but also difficult to predict, quantify, and compare across alternative design choices. A critical yet largely unsupported capability is the early estimation of this effort prior to implementation. Such estimation would enable practitioners to assess feasibility and compare alternative integration options in terms of adaptation cost and complexity. Nevertheless, existing platforms rarely make these cross-view, purpose-dependent dependencies explicit or support reasoning about the effort required to resolve them. As a result, compatibility assessment remains largely expert-driven rather than based on explicit, machine-interpretable modeling constructs~\cite{blair2009models,france2007model,bruel2010viewpoint}.

A key aspect of integration concerns how models interact. \textbf{Integration patterns}—such as one-way data transfer, bidirectional coupling, shared-state coordination, or event-driven interaction—define structured modes of interaction. Each pattern imposes distinct semantic, informational, behavioral, and runtime constraints that directly influence integration feasibility. Consequently, compatibility assessment depends on the chosen integration pattern, as it determines the conditions that must hold across viewpoints. For example, consider coupling a physics model with a biological model of microphytobenthos dynamics, where physics captures heat and radiation processes and biology models nutrient cycles and cell physiology. If a one-way data transfer pattern is used (i.e., the physics model provides outputs such as temperature or radiation that are used as input for the biological model without feedback), integration is relatively straightforward and can be achieved through simple data transformation. However, if a tighter pattern such as bidirectional or integrated coupling is required—where biological processes also influence physical dynamics—additional constraints arise, including synchronization, shared state management, and runtime coordination, making integration significantly more complex. This illustrates how the chosen integration pattern shapes both compatibility conditions and the required integration effort.

Despite their importance, integration patterns are rarely treated as explicit modeling constructs. Instead, these patterns are implicitly embedded within APIs, workflow scripts, and orchestration settings. Support for diverse integration patterns also remains limited, particularly in workflow systems that rely on pipeline-style execution with linear, unidirectional data flow. As a result, integration patterns are often reduced to simple execution order rather than being treated as explicit architectural design choices, hindering systematic and machine-interpretable compatibility assessment.

Addressing these challenges requires treating integration assumptions as explicit modeling elements rather than implicit implementation details. Integration must be approached across multiple architectural dimensions, with semantic, informational, behavioral, and runtime dependencies represented in a structured and analyzable form. This calls for integration-aware metadata that make model semantics, structure, execution behavior, and integration patterns explicit and systematically assessable. Existing standards, however, primarily focus on documentation, execution, or provenance and are typically developed from a single stakeholder or execution perspective. While effective within those contexts, they provide only partial representations and do not capture cross-view, purpose-dependent constraints in a unified way. Consequently, information required to assess reusability and integration feasibility remains implicit or inaccessible to machines. This reveals a structural abstraction gap: DT integration is inherently a multi-viewpoint modeling problem, yet current approaches lack a unified, machine-actionable structure for representing and reasoning about cross-view integration constraints~\cite{ISO10746,bruel2010viewpoint}. We argue that this gap stems from the absence of an explicit, multi-viewpoint modeling abstraction that systematically captures and relates these constraints.

To address this gap, we introduce a multi-viewpoint integration modeling framework grounded in the Reference Model of Open Distributed Processing (RM-ODP)~\cite{ISO10746}. RM-ODP provides a principled foundation for structuring systems across Domain, Information, Computational, Engineering, and Technology viewpoints. Building on this foundation, we formalize integration-relevant constructs and cross-view dependencies as machine-actionable modeling elements, resulting in a viewpoint-structured Model Metamodel that supports model description, discovery, and compatibility reasoning. Rather than replacing existing standards, the framework organizes and reconciles concepts from both general-purpose and domain-specific approaches within a unified multi-viewpoint structure, preserving alignment with established practices while enabling integration-aware reuse. 

The framework enables formal and systematic reasoning about cross-view compatibility through a pattern-aware Mismatch Detector. We distinguish between two complementary forms of compatibility checking. First, deterministic rules capture structural compatibility conditions—such as unit consistency, data format alignment, and pattern-dependent constraints including data-flow direction and synchronization requirements—and are defined at the level of individual viewpoints, enabling systematic, viewpoint-scoped compatibility assessment. Second, many integration constraints involve heterogeneous metadata and implicit semantic assumptions that cannot be fully captured through static checks alone. To address this, we introduce a hybrid detection mechanism that combines deterministic rule generation with structured Large Language Model (LLM)–assisted reasoning. This enables both precise verification of explicit constraints and context-aware interpretation of metadata. The detector systematically identifies semantic, informational, and runtime inconsistencies, as well as specification gaps, with respect to the integration purpose and selected integration pattern, and supports reasoning about the effort required to resolve identified mismatches.

By elevating integration assumptions and constraints to explicit, machine-actionable modeling constructs, the framework moves beyond execution-centric interoperability solutions and enables early-stage, design-time compatibility assessment. This supports more informed integration decisions, including feasibility analysis and estimation of adaptation effort, and paves the way for more automated or semi-automated mismatch resolution. We validate the effectiveness of the approach through expert-guided schema development and an empirical evaluation in environmental modeling scenarios, where heterogeneous models, multi-scale data, and diverse integration patterns provide a representative and challenging case study for evaluating cross-view compatibility and systematic reuse.

This paper makes three main contributions:

\begin{enumerate}

\item \textbf{A Multi-Viewpoint Conceptual Framework for Model Reuse and Integration in DTs.}  
We introduce an RM-ODP–grounded modeling framework that formalizes integration-relevant constructs across complementary viewpoints. The framework reconciles existing standards within a unified structure and makes cross-view dependencies machine-interpretable, providing a principled foundation for interoperability and reuse.

\item \textbf{Viewpoint-Structured Constraint Formalization and Pattern-Aware LLM-Assisted Compatibility Analysis.}  
We formalize integration patterns as first-class constructs and explicitly capture the cross-view constraints they induce. We operationalize compatibility assessment through a hybrid mechanism that combines rule-based constraint generation with LLM-assisted reasoning, enabling systematic identification of inconsistencies and reasoning about resolution effort.

\item \textbf{Demonstration through a Model Catalogue and Realistic DT Use Cases.}  
We instantiate the framework in environmental DT scenarios and demonstrate how explicit modeling of integration constraints enables systematic compatibility assessment and informed reuse decisions in a representative and challenging case study characterized by heterogeneous models, multi-scale data, and diverse integration patterns.

\end{enumerate}

The remainder of the paper is organized as follows. 
Section~\ref{sec:DTIntegrationRequirements} derives integration requirements from a modeling perspective. 
Section~\ref{sec:related work} reviews related work on DT architectures, integration patterns, and model interoperability. 
Section~\ref{sec:conceptual-framework} presents the proposed multi-viewpoint and LLM-assisted approach for model integration.
Section~\ref{sec:case-study-environmental} describes the case study and metamodel instantiation. 
Section~\ref{sec:evaluation} evaluates compatibility reasoning and reuse support. 
Section~\ref{sec:discussion} discusses modeling implications and limitations, and 
Section~\ref{sec:conclusion} concludes the paper.

\section{DT Model Composition and Integration: Modeling Requirements and Challenges}
\label{sec:DTIntegrationRequirements}

Constructing DTs for scientific and engineering applications requires composing heterogeneous computational models, datasets, and services into a coherent and purpose-aligned system. These artifacts are typically developed independently, under different assumptions, abstraction levels, data representations, behavioral semantics, and execution environments. From a system's modeling perspective, DT integration therefore constitutes a cross-model, cross-view consistency problem spanning semantic, informational, behavioral, and runtime abstraction layers.

In current practice, integration remains largely manual, ad hoc, and implementation-driven. Compatibility decisions are often made late in development, after substantial engineering effort has already been invested. This leads to costly redesign cycles, hidden coupling, and limited reusability.

The following requirements are derived from recurring integration challenges reported in the literature and observed in existing DT platforms. For each challenge, we identify the corresponding modeling requirement and explain why addressing it is difficult in heterogeneous DT ecosystems.

\paragraph{\textbf{Model Discovery and Structured Reuse.}}
Identifying suitable models and datasets for a given DT objective remains challenging. Even when components are available, their descriptions are often incomplete, informal, or distributed across repositories~\cite{wu2023comprehensive,tao2022digital}. Models are characterized along heterogeneous dimensions without a unified descriptive structure~\cite{uhlenkamp2022digital}, which limits systematic comparison and reuse. From a modeling perspective, the core issue is that model descriptions function primarily as documentation artifacts rather than as structured, integration-relevant abstractions.
Systematic DT integration therefore requires explicit, structured, and framework-independent model descriptions that capture purpose, assumptions, inputs and outputs with clear data semantics, behavioral interfaces, and execution dependencies in a human- and machine-interpretable form. This is difficult because model knowledge spans multiple abstraction layers—scientific, informational, behavioral, and runtime—typically managed by different stakeholders. Existing standards address isolated aspects such as semantics or execution but rarely their interdependencies. Cross-view relationships, for example how conceptual assumptions constrain data interpretation or runtime behavior, are seldom formalized. The challenge is therefore structural: it requires a multi-viewpoint modeling abstraction capable of making integration-relevant knowledge explicit and analyzable across layers.

\paragraph{\textbf{Compatibility Reasoning and Feasibility Assessment.}}
Selecting reusable components does not guarantee integrability. Components must be compatible with one another, aligned with the intended DT objective, and consistent with the chosen integration pattern. In practice, compatibility assessment is often informal and postponed until implementation~\cite{jones2020characterising,sturm2021creation}, leading to costly discoveries during deployment or runtime. Compatibility reasoning therefore remains largely expert-driven and embedded within implementation logic.
Systematic DT integration requires explicit modeling of compatibility constraints across semantic, informational, behavioral, and runtime dimensions, enabling early feasibility assessment prior to execution and pattern-dependent activation of constraints. This is difficult because compatibility is inherently multi-dimensional: semantic alignment does not ensure data compatibility, and data compatibility does not imply runtime interoperability. Constraints also vary depending on the interaction pattern and often span multiple viewpoints simultaneously. Most platforms encode compatibility procedurally in APIs, configuration files, and orchestration logic. Because of this, it is difficult to analyze compatibility or reason about it during system design. The key modeling challenge is to make these hidden engineering constraints explicit and represent them as formal, analyzable rules so that integration feasibility can be assessed systematically.

\paragraph{\textbf{Real-Time Flexibility and Scalability.}}
DTs frequently require continuous synchronization, real-time coordination, and adaptive interaction. However, many computational models are designed for static or sequential workflows~\cite{lu2020digital,wu2023comprehensive}. Different integration patterns—such as loose, shared, integrated, or embedded coupling—introduce distinct communication semantics, synchronization requirements, and runtime dependencies. Yet these assumptions are typically addressed at the infrastructure level rather than represented explicitly within modeling abstractions.
Systematic DT integration therefore requires treating integration patterns as first-class modeling constructs, with explicit representation of communication, synchronization, and runtime constraints. This is difficult because each pattern activates specific information-level and runtime-level constraints that must be satisfied simultaneously. Tighter coupling strategies may require alignment of execution environments, concurrency semantics, or low-level interoperability mechanisms. These constraints are usually implicit and distributed across configuration artifacts, making them difficult to analyze before deployment. Modeling them requires linking interaction structure, data dependencies, execution semantics, and deployment characteristics within a coherent abstraction. This is inherently a cross-view formalization problem rather than a simple engineering configuration task.

\paragraph{\textbf{Transparency and Quality Management.}}
Reliable DT operation depends on transparency, traceability, validation, and reproducibility~\cite{yao2023systematic}. However, validation status, provenance, uncertainty handling, and licensing conditions are often documented informally or maintained outside the modeling layer. As a result, quality assessment remains largely expert-driven and difficult to automate.
Systematic DT integration requires modeling quality as a first-class cross-view concern, explicitly representing validation status, provenance, assumptions, dependencies, and uncertainty, and linking them through machine-actionable invariants across viewpoints. This is difficult because quality attributes are cross-cutting and viewpoint-dependent. Conceptual validity influences data interpretation and runtime reliability, while legal and technical constraints introduce additional dimensions of quality. Many of these aspects are expressed narratively and lack formal structure, making them challenging to operationalize in automated reasoning. Formalizing quality therefore requires cross-view consistency rules, explicit traceability between abstraction layers, and stronger modeling discipline. In this way, quality becomes a structural integration constraint rather than merely a reporting attribute.

\section{Related Work}
\label{sec:related work}

Surveys consistently show that DT integration challenges arise not merely from data heterogeneity or execution complexity in isolation, but from misalignments across semantic intent, informational structure, behavioral interfaces, and runtime constraints~\cite{tao2022digital,wu2023comprehensive,jones2020characterising}. Furthermore, DT integration spans multiple stakeholder perspectives: domain experts emphasize scientific purpose, data specialists focus on structure and quality, software engineers address interfaces and orchestration, and infrastructure providers concentrate on deployment and performance considerations~\cite{yao2023systematic,trantas2023digital}. 

These observations indicate that DT integration is inherently a multi-viewpoint problem. Based on the requirements analysis in the previous section, systematic reasoning about compatibility and feasibility across these viewpoints is required. In this context, feasibility assessment refers to determining—prior to execution—whether a set of heterogeneous models can be composed consistently with respect to their semantic intent, informational structure, behavioral interfaces, coordination mechanisms, and runtime constraints.

We therefore examine whether existing integration approaches provide sufficient, explicit information to support structured feasibility assessment. For such assessment to be automated, cross-view integration assumptions must be explicitly represented and machine-interpretable rather than embedded implicitly in implementation artifacts.

To evaluate the state of the art, we review three complementary strands of work:

\begin{itemize}
    \item \textbf{Integration Patterns}, to analyze how models interact and identify the types of interaction modes that influence feasibility assessment;
    \item \textbf{Modeling and Integration Frameworks}, to assess whether current platforms explicitly support systematic feasibility reasoning; and
    \item \textbf{Modeling Standards and Metadata}, to determine whether existing representations capture the integration-relevant information required for compatibility analysis.
\end{itemize}

Although these strands have significantly advanced interoperability and execution infrastructure, we demonstrate that they do not provide sufficiently explicit, unified, and machine-actionable integration information to enable systematic feasibility assessment in heterogeneous DT ecosystems.

\subsection{Integration Patterns}

Because feasibility assessment depends on how models are connected and coordinated, we first examine integration pattern research to identify the interaction modes relevant to DT compatibility reasoning. The manner in which models exchange data and synchronize execution introduces constraints related to latency, coupling strength, autonomy, and runtime dependencies. These constraints directly affect whether heterogeneous models can be composed consistently across semantic, informational, behavioral, and execution viewpoints.

Integration patterns categorize these alternative interaction strategies. They describe how computational components exchange data and coordinate execution within a shared environment, spanning a spectrum from loosely coupled, file-based exchange to tightly integrated or embedded configurations.

\begin{figure*}[h]
\centering
\includegraphics[width=0.8\textwidth]{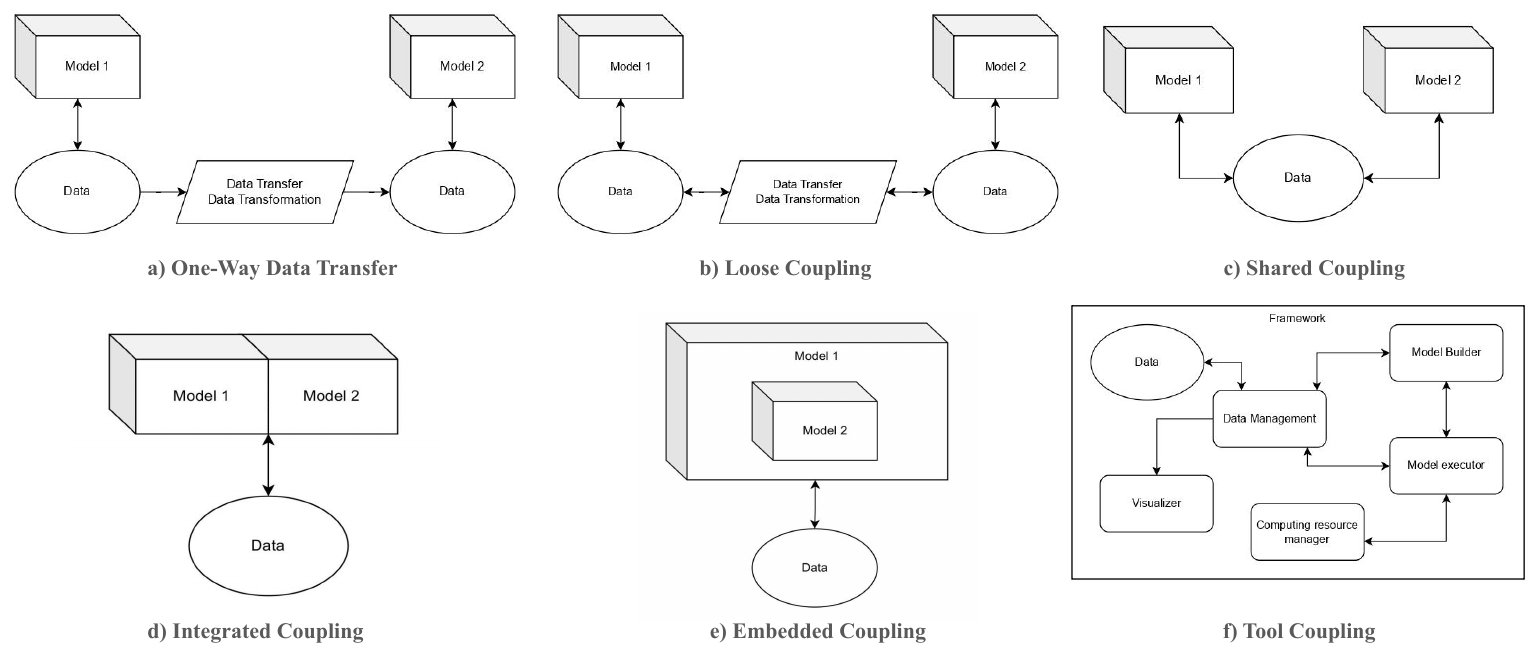}
\caption{Core model integration patterns: (a) \textit{One-Way Data Transfer}, (b) \textit{Loose Integration}, (c) \textit{Shared Integration}, (d) \textit{Integrated Pattern}, (e) \textit{Embedded Pattern}, and (f) \textit{Tool-Orchestrated Integration}.}
\label{fig:IntegrationPatterns}
\end{figure*}

The five-level hierarchy proposed by Brandmeyer and Karimi~\cite{brandmeyer2000coupling} introduced early coupling concepts, primarily addressing static and sequential workflows. However, contemporary DT systems require dynamic reconfiguration, heterogeneous interaction modes, event-driven coordination, and real-time responsiveness. To reflect these additional requirements, we generalize and extend prior classifications into six integration patterns aligned with DT integration needs (Fig.~\ref{fig:IntegrationPatterns}).

\begin{itemize}
    \item \textbf{\textit{One-Way Data Transfer:}} The most decoupled pattern, where one model produces outputs consumed by another without feedback. Communication is typically file-based and asynchronous. This supports batch processing or archival analysis but cannot guarantee continuous synchronization or bidirectional consistency.

    \item \textbf{\textit{Loose Integration:}} Models exchange data bidirectionally via files, APIs, or RPC interfaces while remaining independently deployed. Although more flexible than one-way transfer, this pattern introduces synchronization requirements, interface compatibility constraints, and potential latency overhead.

    \item \textbf{\textit{Shared Integration:}} Models interact through a shared repository or database, enabling frequent or near-real-time exchange. While this improves responsiveness, it imposes concurrency control, shared schema alignment, and state consistency requirements.

    \item \textbf{\textit{Integrated Pattern:}} Models communicate within a shared runtime using in-memory data exchange or direct function invocation. This reduces latency and improves synchronization but increases interdependency and reduces modularity.

    \item \textbf{\textit{Embedded Pattern:}} The tightest coupling form, where one model or service is fully incorporated into another. Execution efficiency is maximized, but autonomy, independent deployment, and reuse flexibility are significantly reduced.

    \item \textbf{\textit{Tool-Orchestrated Integration:}} An orchestration layer dynamically coordinates multiple interaction strategies. This enables heterogeneous ecosystems to combine different coupling modes, but shifts feasibility constraints into orchestration logic and scheduling semantics.
\end{itemize}

Across these patterns, a fundamental distinction arises between \emph{inter-process} and \emph{in-process} interoperability. One-way and loose integration operate across process boundaries, relying on file exchange, APIs, or RPC interfaces. These approaches promote modularity and language independence but introduce serialization overhead, latency, versioning dependencies, and interface alignment challenges~\cite{yang2024multi,nguyen2022programming}.

In contrast, shared, integrated, and embedded patterns rely on in-process interoperability through shared memory, direct invocation, or embedded runtimes. These strategies enable low-latency coordination and tighter synchronization but impose stronger runtime constraints, including \emph{Application Binary Interface (ABI)} compatibility and reliable \emph{Foreign Function Interface (FFI)} mechanisms~\cite{cherny2025metaffi,nguyen2022programming}. As coupling strength increases, performance typically improves while deployment flexibility and autonomy decrease.

From a feasibility assessment perspective, these patterns implicitly activate different informational, behavioral, and runtime constraints. For example, loose integration requires interface compatibility and data format alignment; shared integration requires consistent state management; embedded integration requires ABI-level compatibility and execution co-location. However, existing integration pattern research treats these patterns primarily as architectural design choices or performance trade-offs. The cross-view constraints induced by each pattern are not formalized as explicit modeling constructs, nor represented within a unified, machine-actionable abstraction.

Consequently, although integration patterns clarify how models may interact, they do not provide a structured mechanism for reasoning about whether such interaction is semantically consistent, behaviorally coherent, or operationally feasible prior to execution.

\subsection{Modeling and Integration Frameworks}

After examining interaction patterns as determinants of integration feasibility, we next assess whether existing modeling and integration frameworks provide the explicit information needed to evaluate such feasibility prior to execution. In particular, we ask whether these platforms represent integration assumptions—such as interaction modes, synchronization requirements, data dependencies, and deployment constraints—in a structured and machine-interpretable manner.

Modeling and integration frameworks provide infrastructure for composing and executing computational models. They can be broadly categorized into component-based frameworks, scientific workflow systems, and VREs. While these platforms enable interoperability and coordinated execution, their primary focus is operational orchestration rather than formal representation of cross-view integration constraints.

From a feasibility assessment perspective, the key issue is whether integration-relevant assumptions are made explicit. If assumptions about interfaces, coordination semantics, data alignment, and runtime compatibility remain embedded within APIs, workflow definitions, middleware configurations, or execution scripts, compatibility reasoning cannot be performed systematically without executing the system or manually inspecting implementation details.

\begin{itemize}

\item \textbf{Component-Based Frameworks:}
Component-based frameworks such as OpenMI~\cite{gregersen2007openmi}, ESMF~\cite{hill2004esmf}, and CSDMS~\cite{peckham2013csdms} encapsulate models as interoperable components with standardized interfaces. These frameworks enable efficient coupling within specific scientific domains, including environmental modeling. However, they typically embed domain-specific assumptions, data structures, and rigid interface definitions. As a result, reuse across heterogeneous contexts requires substantial adaptation, and compatibility is assessed primarily at the interface level rather than through explicit modeling of semantic and behavioral constraints.

\item \textbf{Scientific Workflow Systems:}
Scientific workflow systems such as Kepler~\cite{altintas2004kepler}, Taverna~\cite{oinn2004taverna}, and VisTrails~\cite{callahan2006vistrails} provide graphical environments for composing models, datasets, and analytical tools, often with provenance tracking and reproducibility support. However, their execution semantics are typically pipeline-oriented and static, limiting support for asynchronous, event-driven, or adaptive interaction modes required by real-time DT systems. Integration decisions are encoded within workflow structure and execution logic, rather than represented as explicit, reusable compatibility constraints.

\item \textbf{Virtual Research Environments (VREs):}
VREs such as myExperiment~\cite{rojas2011myexperiment}, Galaxy~\cite{goecks2010galaxy}, and NaaVRE~\cite{zhao2022notebook} extend workflow systems with collaborative repositories, execution backends, and user-facing services. These platforms facilitate discovery and sharing of models and data. Nevertheless, the metadata they maintain is often descriptive and informal, lacking structured representations of behavioral dependencies, interaction assumptions, and deployment constraints required for automated compatibility reasoning in DT ecosystems.

\end{itemize}

Across these frameworks, integration assumptions remain implicit within implementation artifacts rather than explicitly modeled. Although these platforms are valuable as execution infrastructure, they do not provide a modeling abstraction that systematically separates and relates semantic intent, informational structure, behavioral interfaces, coordination semantics, and runtime constraints. As a result, compatibility reasoning, reuse assessment, and mismatch detection remain largely manual and context-dependent.

\subsection{Modeling Standards and Metadata}
Since feasibility assessment requires explicit, machine-actionable representations of cross-view integration assumptions, we next examine whether existing modeling standards and metadata initiatives provide the necessary information for such assessment. Existing standards, however, address these concerns in isolation and from limited stakeholder perspectives.

Despite these limitations, existing initiatives provide valuable building blocks and can be broadly divided into domain-specific and general-purpose standards. Using environmental modeling as a representative use case, we highlight their respective strengths and limitations.

\begin{itemize}

\item \textbf{Domain-Specific Modeling Standards:}
Standards widely used in environmental and scientific modeling span different abstraction levels. At the conceptual level, frameworks such as ODD~\cite{grimm2020odd} and TRACE~\cite{grimm2014towards} provide structured narrative descriptions of model purpose, assumptions, processes, and evaluation practices, thereby improving transparency and reproducibility. However, they rely primarily on textual documentation and lack formal, machine-actionable representations suitable for automated discovery, semantic alignment, or compatibility reasoning. At the execution level, standards such as SED-ML~\cite{waltemath2011reproducible} and SBML Level~3~\cite{keating2020sbml} encode model structure and simulation protocols in machine-readable formats, enabling reproducible execution and syntactic interoperability. While effective for simulation exchange, they focus predominantly on execution concerns and provide limited support for representing higher-level semantic intent, interaction assumptions, or cross-model integration constraints required for heterogeneous DT ecosystems.

\item \textbf{General-Purpose Metadata Initiatives:}
Initiatives such as FAIR4RS~\cite{chue2022fair} and the Open Modeling Foundation (OMF)~\cite{omf} promote principles of transparency, interoperability, and reuse across disciplines. Although influential, they remain intentionally abstract and do not define concrete modeling constructs for capturing integration patterns, behavioral dependencies, coordination semantics, or deployment constraints required for systematic compatibility reasoning.

\end{itemize}

Taken together, these standards provide partial and fragmented views of models—semantic documentation, execution encoding, or high-level reuse principles—but do not offer a unified, machine-actionable modeling representation that captures and relates semantic, informational, behavioral, coordination, and deployment concerns. Moreover, each is typically developed from the perspective of a specific stakeholder community, leading to implicit and distributed representations of integration-relevant assumptions.

\subsection{Gap Analysis}
\label{subsec:gap}

The reviewed strands—Integration Patterns, Modeling and Integration Frameworks, and Modeling Standards and Metadata—contribute important insights toward DT interoperability. However, when evaluated through the lens of feasibility assessment, they remain fragmented and insufficient for systematic, pattern-aware compatibility reasoning.

Integration pattern research clarifies the architectural alternatives through which models interact, ranging from loosely coupled data exchange to tightly embedded execution. It identifies performance trade-offs and coupling characteristics associated with each pattern. Yet these patterns are treated primarily as design choices rather than as sources of explicit integration constraints. The semantic, informational, behavioral, coordination, and runtime implications induced by different interaction modes are not formalized as structured modeling constructs. As a result, while patterns describe how models are connected, they do not provide a principled mechanism for reasoning about whether those connections are semantically consistent, behaviorally coherent, or operationally feasible across viewpoints.

Modeling and integration frameworks implement these interaction strategies through execution infrastructure. However, the integration assumptions inherent in these patterns—such as synchronization requirements, interface contracts, coordination semantics, and deployment dependencies—as well as other information required for feasibility assessment, including semantic alignment, are not represented explicitly. Instead, they remain embedded implicitly within APIs, workflow definitions, middleware configurations, and orchestration logic. Because these constraints are encoded procedurally rather than represented declaratively as modeling constructs, they cannot be systematically analyzed prior to execution. As a result, compatibility assessment depends on manual inspection, trial deployment, or runtime experimentation rather than structured, model-based reasoning.

Modeling standards and metadata initiatives provide formal descriptions of selected aspects of models, such as semantic documentation or executable structure. However, these representations address concerns in isolation and do not relate semantic intent, informational structure, behavioral interfaces, coordination semantics, and runtime constraints within a unified abstraction. Cross-view dependencies remain distributed across heterogeneous artifacts and stakeholder perspectives, preventing comprehensive feasibility analysis.

Overall, DT integration is inherently multi-viewpoint and interaction-pattern-dependent, yet existing approaches lack an explicit, unified representation of the cross-view constraints activated by different interaction modes. Consequently, compatibility assessment, reuse evaluation, and mismatch detection remain largely informal, manual, and context-specific. This reveals a structural modeling gap: there is no formal, machine-actionable abstraction that treats integration patterns as first-class architectural elements and systematically captures the semantic, informational, behavioral, coordination, and runtime constraints they induce. To address this gap, the next section introduces a multi-viewpoint integration modeling framework. The framework elevates integration patterns to explicit architectural constructs and formalizes cross-view dependencies as machine-actionable constraints, enabling systematic feasibility assessment and structured compatibility reasoning in heterogeneous DT ecosystems.

\section{Multi-Viewpoint Integration Framework and LLM-Assisted Compatibility Analysis}
\label{sec:conceptual-framework}

This section presents a multi-viewpoint--based and LLM-assisted approach for systematic model integration in DT ecosystems. As discussed in Section~\ref{subsec:gap}, existing solutions rarely make the cross-view constraints induced by integration patterns explicit and machine-actionable: integration assumptions remain embedded in implementation artifacts, patterns are treated as architectural styles rather than constraint sources, and feasibility assessment often relies on manual reasoning or runtime experimentation.

To address this gap, we introduce (i) a unified viewpoint-structured metadata schema for describing reusable DT models and intended integrations, and (ii) pattern-aware integration services that derive explicit compatibility constraints and assess them using structured LLM-based reasoning. The approach adds a declarative modeling layer that can be attached to existing modeling and integration platforms to enable structured compatibility assessment prior to deployment.

\begin{figure*}[h]
\centering
\includegraphics[width=0.9\columnwidth]{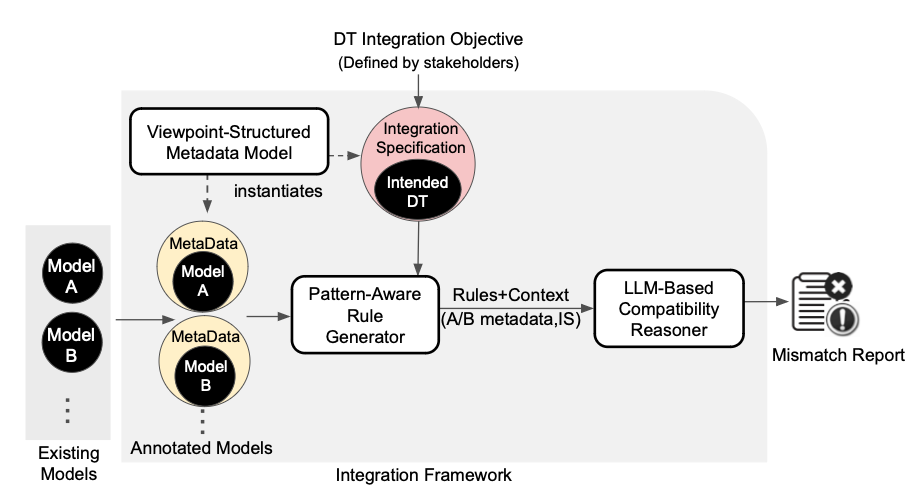}
\caption{Conceptual workflow of the proposed multi-viewpoint integration approach.}
\label{fig:conceptual_framework}
\end{figure*}

Fig.~\ref{fig:conceptual_framework} summarizes the workflow. Stakeholders specify a \emph{DT integration objective}, which is captured as an \emph{Integration Specification (IS)}. The IS constitutes a partial instantiation of the viewpoint-structured metadata model and records the intended DT configuration at a requirement level, including the integration goal and selected integration pattern. In parallel, candidate reusable models are described using the same metadata schema, yielding machine-interpretable profiles across the RM-ODP viewpoints. Based on these descriptions, the \emph{Pattern-Aware Rule Generator} derives explicit compatibility constraints, and the \emph{LLM-Assisted Compatibility Reasoner} evaluates them to produce a structured mismatch report that distinguishes satisfied conditions from violated requirements and specification gaps, supporting feasibility assessment before implementation.

The following subsections detail the underlying methodology and the design of the two core components: the viewpoint-structured Metadata Schema (Section~\ref{subsec:metamodel}) and the Pattern-Aware and LLM-Assisted Integration Services (Section~\ref{subsec:mismatch-detection}). The section concludes with the formal metamodel that underpins the integration framework (Section~\ref{subsec:FormalFoundation}).

\subsection{Design and Structuring of the RM-ODP--Based Metadata Schema}
\label{subsec:metamodel}

The viewpoint-structured metadata schema is the first core artifact of the proposed approach. It provides an integration-oriented representation of DT models and intended integrations in a unified, machine-interpretable form. While existing standards capture selected aspects of model description, they typically do not relate semantic intent, data characteristics, computational interfaces, coordination semantics, and execution constraints within a single modeling abstraction. The proposed schema addresses this limitation by organizing integration-relevant information along complementary architectural dimensions and by making cross-view dependencies explicit.

\subsubsection{Viewpoint-Based Structuring Principle}

\begin{figure*}[h!]
\centering
\includegraphics[width=0.8\textwidth]{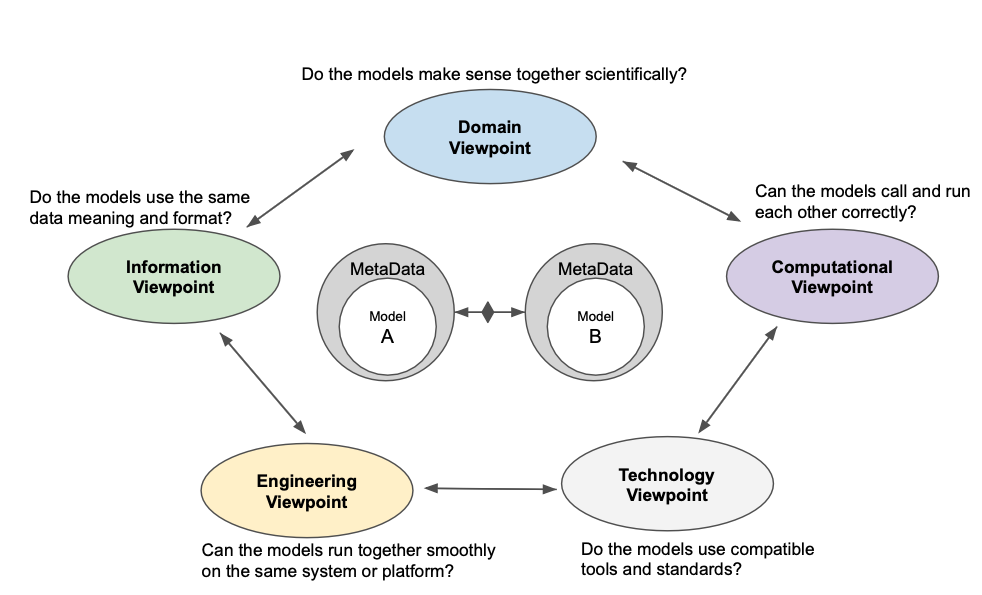}
\caption{RM-ODP viewpoints as structuring dimensions for integration-relevant metadata in DT ecosystems.}
\label{fig:RM-ODPviewpoints}
\end{figure*}

The schema is organized according to the five RM-ODP viewpoints: \textit{Domain (Enterprise)}, \textit{Information}, \textit{Computational}, \textit{Engineering}, and \textit{Technology}. Each viewpoint isolates a distinct class of integration concerns while enabling traceable relations across abstraction layers (Fig.~\ref{fig:RM-ODPviewpoints}).

\begin{enumerate}
\item \textbf{\textit{Domain (Enterprise)} viewpoint:} Captures the model purpose, conceptual scope, methodological assumptions, intended use cases, and domain constraints, establishing semantic and conceptual alignment.
\item \textbf{\textit{Information} viewpoint:} Defines the structure and semantics of model data, including variables, units, ranges, types, spatial/temporal coverage, coordinate systems, and provenance, enabling data-level compatibility assessment.
\item \textbf{\textit{Computational} viewpoint:} Specifies functional behavior and interaction interfaces (e.g., APIs, signatures, I/O bindings, events, error handling, abstract workflow steps), supporting compositional interaction and orchestration.
\item \textbf{\textit{Engineering} viewpoint:} Describes distributed execution and coordination (e.g., orchestration logic, communication mechanisms, synchronization, runtime dependencies, scalability constraints).
\item \textbf{\textit{Technology} viewpoint:} Records concrete implementation details (e.g., language, runtime environment, container images, hardware needs, file formats, serialization, licensing, reproducibility artifacts).
\end{enumerate}

This structuring makes it possible to express integration requirements and assess compatibility in a principled way: each viewpoint can be checked independently while cross-view constraints remain traceable through the shared schema.

\subsubsection{Schema Engineering Process}
\label{SchemaEngineeringProcess}

\begin{figure*}[h!]
\centering
\includegraphics[width=\textwidth]{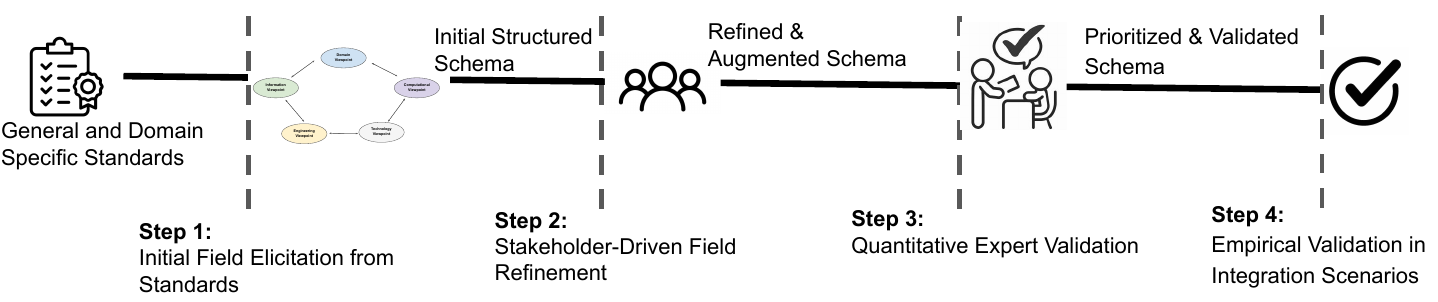}
\caption{RM-ODP--guided metadata schema design process.}
\label{fig:schema_engineering_process}
\end{figure*}

Concrete metadata elements were derived through a four-phase design and validation process (Fig.~\ref{fig:schema_engineering_process}). RM-ODP provides the conceptual scaffold, while standards analysis and stakeholder input shape the schema content and requiredness.

\textbf{Phase 1 -- Initial field elicitation from standards.}
We analyzed general-purpose and domain-specific standards for model description, interoperability, and metadata management. Extracted fields were mapped to RM-ODP viewpoints, yielding an initial inventory aligned with abstraction layers.

\textbf{Phase 2 -- Stakeholder-driven field refinement.}
Semi-structured interviews with cross-domain DT roles (domain experts, data/semantic specialists, software developers, infrastructure managers, and integration specialists) identified integration-relevant attributes not captured in standards, clarified implicit assumptions, and surfaced cross-layer dependencies. The schema was refined via viewpoint-level augmentation and semantic clarification to remain domain-agnostic while reflecting practical integration needs.

\textbf{Phase 3 -- Quantitative expert validation.}
A survey-based assessment asked experts to rank metadata elements by importance for integration and reuse. Results informed viewpoint-level completeness and supported classification of required versus optional fields.

\textbf{Phase 4 -- Empirical validation in integration scenarios.}
The schema was instantiated in environmental DT use cases. Observed mismatch patterns and integration outcomes provided feedback on adequacy and expressiveness and informed iterative refinements.

\textbf{Example 1.} To illustrate instantiation without introducing the full empirical study, Table~\ref{tab:metadata_instantiation} presents a partial viewpoint-structured description of two representative environmental simulation models (FLake and PCLake+) together with an intended IS defined under a One-Way data-transfer pattern. The table includes only a representative subset of metadata elements to convey the structure and organization of the schema; the complete set of extracted and validated metadata fields is detailed in Section \ref{sec:case-study-environmental}. This example highlights how heterogeneous models and an integration objective are represented consistently across viewpoints using the proposed schema.

\begin{table*}[t]
\centering
\caption{Illustrative (partial) instantiation of the viewpoint-structured metadata model for two reusable models and an intended IS.}
\label{tab:metadata_instantiation}
\footnotesize
\begin{tabular}{p{0.13\linewidth}p{0.26\linewidth}p{0.26\linewidth}p{0.2\linewidth}}
\hline
\textbf{RM-ODP Viewpoint} & \textbf{Model A: FLake} & \textbf{Model B: PCLake+} & \textbf{Intended IS} \\
\hline

\textbf{Domain} &
Purpose: Thermal stratification simulation \newline
Scope: Freshwater temperate lakes \newline
Model Type: Mechanistic \newline
Assumptions: 1D vertical column model &
Purpose: water quality and trophic states \newline
Scope: freshwater lakes \newline
Model Type: Process-based ecological \newline
Assumptions: 0D, epilimnion and hypolimnion layer-averaged representation &
Purpose: Climate--ecology--ecosystem service assessment \newline
Pattern: One-Way integration \newline
Goal: Use physical outputs to drive ecology \\

\textbf{Information} &
Output: Water temperature ($^\circ$C), Mixing depth (m) \newline
Dimensionality: 1D vertical profile \newline
Time Steps: Hourly \newline
Spatial Resolution: Depth-resolved \newline
Format: NetCDF &
Input datasets: Water temperature (K), Mixing depth (m), Flow rate (mm/day), nutrient loading (mg/m2/day) \newline
Dimensionality: Scalar layer-averaged \newline
Time Steps: Daily \newline
Spatial Resolution: Whole-lake average \newline
Format: CSV &
Required variables: Climate data, Mixing depth, Water balance, Nutrient loading, Lake characteristics \newline
Expected time step: Daily \newline
Layer-averaged scalar inputs \\

\textbf{Computational} &
Interface signature: Batch file export \newline
Error handling: Log-based reporting &
Interface signature: CSV batch import \newline
Error handling: Execution abort on format mismatch &
Pattern: One-Way (A $\rightarrow$ B) \newline
Sequential invocation assumed \\

\textbf{Engineering} &
Execution constraints: Offline batch processing \newline
Parallel execution: No &
Execution constraints: Scheduled daily execution \newline
Parallel execution: Limited &
Assumes batch-based coordination \\

\textbf{Technology} &
Language: Fortran \newline
OS: Linux \newline
Software specification: Docker container \newline
License: MIT &
Language: C++ \newline
OS: Windows/Linux \newline
Software specification: Native executable \newline
License: Apache 2.0 &
No additional constraints specified \\
\hline
\end{tabular}
\end{table*}

\subsection{Pattern-Aware and LLM-Assisted Integration Services}
\label{subsec:mismatch-detection}

Compatibility reasoning is operationalized through a two-stage hybrid mismatch detection engine (Fig.~\ref{fig:conceptual_framework}). First, a \emph{Pattern-Aware Rule Generator} deterministically derives viewpoint-scoped compatibility rules from the IS, the selected integration pattern, and the activated data-flow dependencies. Second, an \emph{LLM-Based Compatibility Reasoner} evaluates each instantiated rule using the relevant metadata context (models $A$, $B$, and the IS) and records a verdict (\emph{Match}, \emph{Mismatch}, or \emph{Gap}) together with an actionable recommendation in a structured mismatch report.

\subsubsection{Mismatch Taxonomy}

Integration mismatches are classified into two categories:
\begin{itemize}
\item \textbf{Pattern-agnostic mismatches:}
Incompatibilities instantiated for every integration attempt, independent of the integration pattern. These include conceptual misalignment, inconsistent domain assumptions, missing provenance, and other foundational metadata inconsistencies that establish baseline requirements.
\item \textbf{Pattern-specific mismatches:}
Incompatibilities activated by the selected integration pattern. These include (i) \emph{information-level mismatches} induced by data-flow dependencies (e.g., schema, unit, spatial/temporal resolution incompatibilities), and (ii) \emph{runtime-level mismatches} induced by tighter interaction semantics (e.g., synchronization, coupling, concurrency, ABI or execution-environment constraints).
\end{itemize}
This distinction reflects that compatibility depends not only on individual models but also on the interaction semantics imposed by the integration pattern.

\subsubsection{Formal Constraint Specification}

The detection mechanism is grounded in a constraint model derived from the integration metamodel (Section~\ref{subsec:FormalFoundation}). A compatibility constraint is defined as:
\[
c = \langle scope, category, expression \rangle,
\]
where $scope$ denotes the RM-ODP viewpoint to which the constraint applies,
$category \in \{\textit{PatternAgnostic}, \textit{InformationLevel}, \textit{RuntimeLevel}\}$,
and $expression$ specifies a compatibility condition over metadata instances and, when applicable, activated data-flow edges.

Constraints are instantiated deterministically from the IS and pattern semantics and are evaluated by the LLM-based reasoner. A \emph{Mismatch} verdict indicates that the compatibility requirement is violated; a \emph{Gap} indicates missing or insufficient metadata to assess the condition. Both outcomes are recorded as structured report entries linked to the corresponding IS, together with an explanation and a repair suggestion.

\subsubsection{Hybrid Detection Procedure}

Algorithm~\ref{alg:pattern-aware-detector-merged} summarizes the hybrid detection procedure. Let $(A,B)$ denote an ordered pair of annotated models selected from $M$, and let $E_p^{AB}$ denote the set of directed data-flow edges activated between $A$ and $B$ under the integration pattern $p$.

\begin{algorithm}[h]
\footnotesize
\caption{Pattern-aware and LLM-assisted mismatch detection}
\label{alg:pattern-aware-detector-merged}
\begin{algorithmic}[1]
\Require Annotated models $M = \{m_1, \dots, m_n\}$ with multi-viewpoint metadata (Section~\ref{subsec:metamodel}); IntegrationSpecification $IS$
\Ensure $\mathcal{B}$: Structured mismatch report entries (Mismatch/Gap)

\State $\mathcal{B} \gets \emptyset$
\State $p \gets IS.\text{IntegrationPattern}$

\ForAll{ordered pairs $(A,B)$ where $A,B \in M$}
\Statex \rule{\linewidth}{0.4pt}

\Statex \textbf{Stage 1: Pattern-Aware Rule Generation}
\State $\Phi \gets \emptyset$

\State $\Phi \gets \Phi \cup$ \textsc{GeneralRules}$(A,B,IS)$ \Comment{Pattern-agnostic rules}

\State $deps_{A\to B} \gets$ \textsc{Consumes}$(B, A.\text{Output})$
\State $deps_{B\to A} \gets$ \textsc{Consumes}$(A, B.\text{Output})$

\If{$p = \text{One-Way}$}
    \State $E_{p}^{AB} \gets \{A \rightarrow B \mid deps_{A\to B}\}$
\ElsIf{$p \in \{\text{Loose, Shared}\}$}
    \State $E_{p}^{AB} \gets \{A \rightarrow B \mid deps_{A\to B}\} \cup \{B \rightarrow A \mid deps_{B\to A}\}$
\ElsIf{$p \in \{\text{Integrated, Embedded}\}$}
    \State $E_{p}^{AB} \gets \{A \rightarrow B \mid deps_{A\to B} \land IS.\text{enables}(A\to B)\} \cup$
    \Statex \hspace*{2em} $\{B \rightarrow A \mid deps_{B\to A} \land IS.\text{enables}(B\to A)\}$
\EndIf

\ForAll{$e \in E_{p}^{AB}$}
    \State $\Phi \gets \Phi \cup$ \textsc{InfoRules}$(e, IS)$ \Comment{information-level rules per active edge}
\EndFor

\State $\Phi \gets \Phi \cup$ \textsc{RuntimeRules}$(A,B,IS,p)$ \Comment{runtime-level rules induced by $p$}

\Statex \rule{\linewidth}{0.4pt}

\Statex \textbf{Stage 2: LLM-Based Compatibility Reasoner}
\State $T \gets$ \textsc{PromptTemplate}$(p,\Phi)$ \Comment{instructions + output JSON schema}
\ForAll{instantiated rules $r \in \Phi$}
    \State $ctx_r \gets$ \textsc{SelectContext}$(r, A,B,IS)$ \Comment{only metadata relevant to $r$}
    \State $o \gets$ \textsc{LLM\_Evaluate}$(T, r, ctx_r)$
    \State $\hat{o} \gets$ \textsc{ValidateAndParseJSON}$(o)$ \Comment{schema check; retry/fallback on failure}
    \If{$\hat{o}.\text{verdict} \in \{\text{Mismatch}, \text{Gap}\}$}
        \State $\mathcal{B} \gets \mathcal{B} \cup \{\textsc{Report}(r, \hat{o})\}$ \Comment{includes justification and suggested mediation}
    \EndIf
\EndFor

\EndFor
\State \Return $\mathcal{B}$
\end{algorithmic}
\end{algorithm}





\textbf{Stage 1 -- Pattern-aware rule generation.}
For each annotated model pair $(A,B)$, the rule generator derives a set of compatibility constraints $\Phi$ from the Integration Specification (IS) and the selected integration pattern $p$. These constraints include (i) baseline rules that apply independently of the integration pattern and (ii) additional information- and runtime-level rules induced by the active interaction edges $E_p^{AB}$.

\textbf{Stage 2 -- LLM-assisted compatibility reasoning.}
Each instantiated constraint $r \in \Phi$ is evaluated using a structured prompt that provides the rule and the relevant metadata context (the fields referenced by $r$ from the annotated models $A$, $B$, and the IS). The LLM-based reasoner returns a compatibility verdict (\emph{Match}, \emph{Mismatch}, or \emph{Gap}) together with an explanation and, when applicable, a suggested resolution strategy. The resulting outputs are aggregated into a structured mismatch report that supports integration analysis and design refinement prior to implementation.

\textbf{Example 2.}
Building on the metadata instantiation in Example~1 (Table~\ref{tab:metadata_instantiation}), 
Table~\ref{tab:field_level_results_compact} shows the mismatch report generated by the LLM-assisted module. It evaluates the pattern-aware rules derived from the FLake $\rightarrow$ PCLake+ One-Way IS and produces a structured compatibility report, including a verdict (\emph{Match}, \emph{Mismatch}, or \emph{Gap}) and an actionable recommendation for each field.

\begin{table}[t]
\centering
\footnotesize
\caption{Illustrative field-level compatibility assessment for FLake $\rightarrow$ PCLake+ under a One-Way integration pattern.}
\label{tab:field_level_results_compact}
\begin{tabular}{p{0.22\linewidth}p{0.12\linewidth}p{0.58\linewidth}}
\hline
\textbf{Field} & \textbf{Verdict} & \textbf{Recommended Adaptation} \\
\hline
\multicolumn{3}{l}{\textbf{Domain Viewpoint}} \\
Purpose & Match & No conceptual conflict detected between physical and ecological model roles. \\
Scope & Gap & Verify that both models target compatible lake conditions and usage context. \\
Assumptions & Mismatch & Introduce vertical aggregation to convert the 1D temperature profile into a layer-averaged scalar input. \\
\hline
\multicolumn{3}{l}{\textbf{Information Viewpoint}} \\
Temperature Unit & Mismatch & Apply unit conversion ($^\circ$C $\rightarrow$ K) prior to data exchange. \\
Dimensionality & Mismatch & Aggregate the depth-resolved temperature profile to a layer-averaged scalar representation. \\
Temporal Resolution & Mismatch & Resample hourly outputs to daily values before passing data to Model B. \\
Data Format & Mismatch & Implement a NetCDF $\rightarrow$ CSV transformation pipeline. \\
Mixing Depth & Match & Variable is available in Model A and accepted as input by Model B. \\
\hline
\multicolumn{3}{l}{\textbf{Computational Viewpoint}} \\
Interface Signature & Match & Batch-based file exchange is compatible with the One-Way integration pattern. \\
Error Handling & Gap & Introduce input validation or preprocessing to prevent runtime failures due to format mismatches. \\
\hline
\multicolumn{3}{l}{\textbf{Engineering Viewpoint}} \\
Execution Constraints & Match & Schedule Model A execution before Model B within the integration workflow. \\
\hline
\multicolumn{3}{l}{\textbf{Technology Viewpoint}} \\
Operating System & Match & Ensure deployment in a Linux-compatible execution environment. \\
License & Match & Both models use permissive open-source licenses that allow integration and redistribution. \\
\hline
\end{tabular}
\end{table}

\subsection{Formal Foundation of the Integration Framework}
\label{subsec:FormalFoundation}

\begin{figure*}[h!]
\centering
\includegraphics[width=0.8\textwidth]{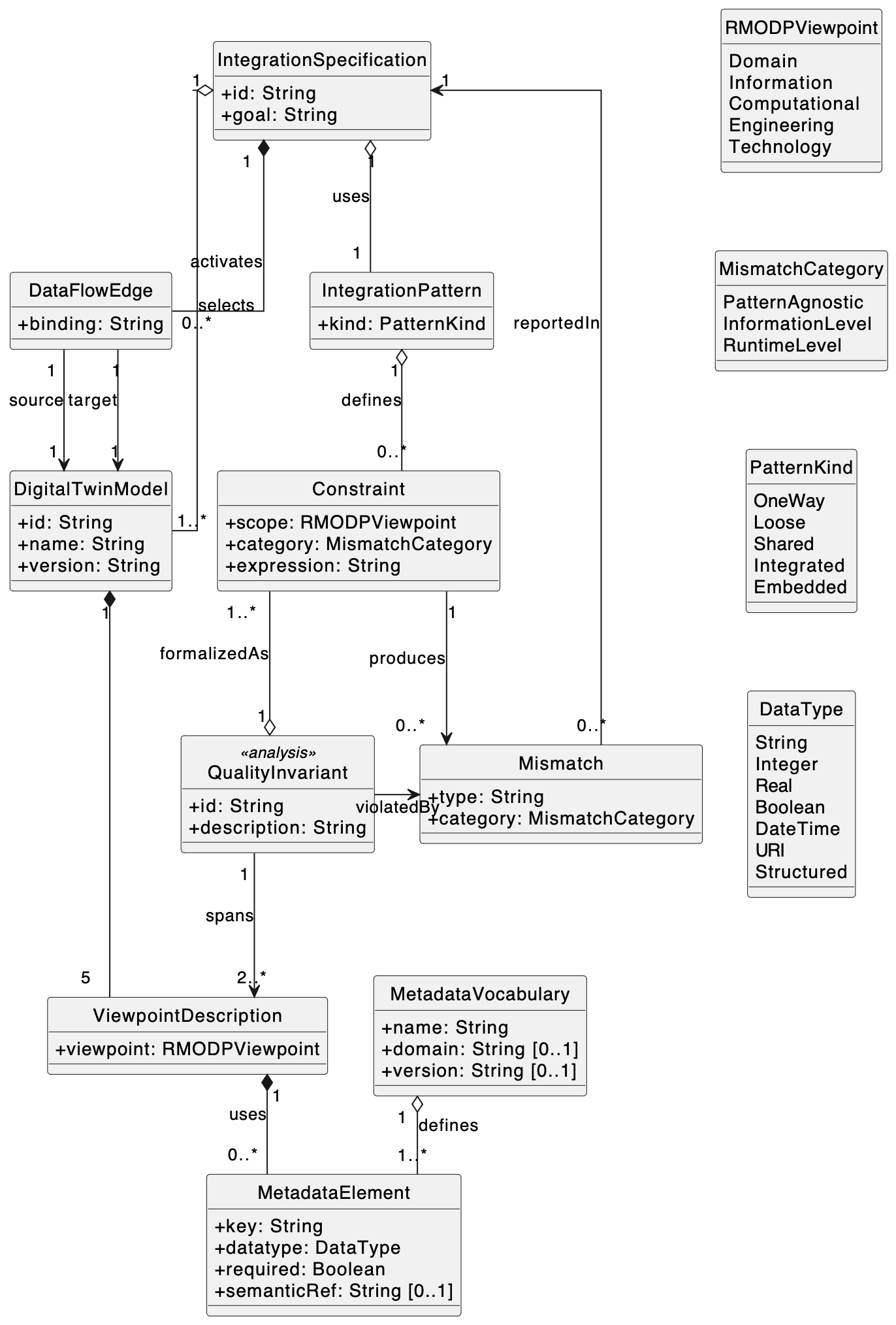}
\caption{Formal architecture of the multi-viewpoint integration framework, defining the structural elements and relationships supporting model description, integration configuration, and constraint-based reasoning.}
\label{fig:integration_metamodel}
\end{figure*}

To enable systematic reasoning and tool support, the proposed approach is formalized as an integration metamodel that defines the structural elements and relationships governing model description, integration configuration, and constraint-based reasoning (Fig.~\ref{fig:integration_metamodel}). The metamodel makes explicit how integration intent, model metadata, and compatibility assessment are formally connected.

At the modeling level, a \textit{DigitalTwinModel} is composed of five \textit{ViewpointDescriptions}, each aggregating structured \textit{MetadataElements}. Metadata elements are defined with respect to a \textit{MetadataVocabulary}, enabling controlled extensibility while preserving formal consistency.

At the integration level, an \textit{IntegrationSpecification} selects one or more models and associates them with a single \textit{IntegrationPattern}. The selected pattern activates directed \textit{DataFlowEdges}, which represent structural information dependencies between models.

At the reasoning level, compatibility is expressed through viewpoint-scoped \textit{Constraints}, categorized as \textit{PatternAgnostic}, \textit{InformationLevel}, or \textit{RuntimeLevel}. Constraints are instantiated over metadata elements and, when applicable, activated data-flow edges. Their evaluation produces structured \textit{Mismatch} instances linked to the corresponding IS, with outcomes classified as \textit{Match}, \textit{Mismatch}, or \textit{Gap}.

In addition, \textit{QualityInvariant} constructs capture cross-view consistency requirements by formalizing higher-level properties through sets of constraints. Violations are reflected as derived compatibility issues.

Through this formalization, the framework integrates model description, integration intent, structural dependency activation, and constraint-based reasoning within a single coherent architecture. The metamodel therefore provides the formal foundation that enables systematic, pattern-aware compatibility assessment prior to implementation.

\section{Case Study: Applying the Model Integration Framework to Environmental Modeling}
\label{sec:case-study-environmental}

To demonstrate how the framework introduced in Section~\ref{sec:conceptual-framework} can be instantiated in a concrete setting, we apply it to the environmental modeling domain. Environmental DT applications commonly combine heterogeneous simulation and data-driven models, operate over diverse spatiotemporal scales, and must interoperate across disciplinary and organizational boundaries. These characteristics make the domain a demanding setting for exercising both the expressiveness of the RM-ODP–guided metadata schema and the applicability of the mismatch-detection logic.

This case study focuses on \emph{instantiation}: we (i) specialize the generic metadata schema with environmental semantics while preserving its five-viewpoint structure, and (ii) derive a domain-specific mismatch taxonomy by binding the generic constraint categories to the instantiated metadata fields. The instantiation was developed in collaboration with environmental modeling experts to ensure that the selected metadata elements and integration assumptions reflect realistic modeling and reuse practices. The evaluation of the resulting schema and mismatch detector is reported separately in Section~\ref{sec:evaluation}.

\subsection{Metadata Schema Instantiation for Environmental Models}
\label{subsec:Metamodel-Environmental-Models}

Following the RM-ODP–based structuring principles introduced in Section~\ref{subsec:metamodel} and the four-phase schema engineering process described in Section~\ref{SchemaEngineeringProcess}, the generic model metadata schema was instantiated for the environmental modeling domain.

\textbf{Phase 1 – Standards-Based Field Elicitation.}
Representative modeling and metadata standards relevant to environmental modeling were systematically analyzed to extract existing metadata elements and assess their coverage across the five RM-ODP viewpoints. A total of 6 standards were examined. Table~\ref{tab:standards_viewpoints_roles} summarizes their coverage of integration concerns and stakeholder roles. The analysis revealed that existing standards typically address only subsets of viewpoints and are tailored to particular stakeholder communities, resulting in fragmented representation of integration-relevant aspects.
\begin{table*}[t]
\centering
\caption{Coverage of RM-ODP viewpoints and stakeholder roles by representative modeling and metadata standards.
Symbols: ``--'' = no coverage, ``(\checkmark)'' = partial coverage, ``\checkmark'' = full coverage.}
\label{tab:standards_viewpoints_roles}
\footnotesize
\setlength{\tabcolsep}{4pt}
\renewcommand{\arraystretch}{1.05}

\begin{tabular}{
p{0.16\linewidth}
p{0.08\linewidth}
p{0.06\linewidth}
p{0.06\linewidth}
p{0.06\linewidth}
p{0.06\linewidth}
p{0.4\linewidth}}
\hline
\textbf{Standard} &
\textbf{Domain} &
\textbf{Info} &
\textbf{Comp.} &
\textbf{Eng.} &
\textbf{Tech.} &
\textbf{Roles addressed $^{*}$} \\
\hline

ODD~\cite{grimm2020odd} &
\checkmark &
(\checkmark) &
-- &
-- &
(\checkmark) &
Domain Scientists, Measurement Model Designer, Semantic Curator, Users \\
\hline

TRACE~\cite{grimm2014towards} &
\checkmark &
(\checkmark) &
(\checkmark) &
(\checkmark) &
(\checkmark) &
Domain Scientists, Data Curator, Semantic Curator, Users \\
\hline

SED-ML~\cite{waltemath2011reproducible} &
(\checkmark) &
\checkmark &
-- &
(\checkmark) &
(\checkmark) &
Developers, Data Scientists, Users, Infrastructure Managers \\
\hline

SBML~\cite{keating2020sbml} &
(\checkmark) &
\checkmark &
-- &
(\checkmark) &
(\checkmark) &
Developers, Data Scientists, Users \\
\hline

FAIR4RS~\cite{chue2022fair}&
(\checkmark) &
(\checkmark) &
-- &
-- &
(\checkmark) &
Asset Providers, FAIRification Stewards, Data Stewards, Funding Organizations \\
\hline

OMF~\cite{omf} &
(\checkmark) &
(\checkmark) &
-- &
-- &
(\checkmark) &
Partners and Collaborators, Policymakers, Asset Providers, Infrastructure Managers \\
\hline

\hline
RM-ODP–based framework &
\checkmark &
\checkmark &
\checkmark &
\checkmark &
\checkmark &
Domain Scientists, Data Curators, Semantic Curators, Software Developers, Infrastructure Managers, DT Integrators, Platform Operators, Users \\
\hline

\end{tabular}
$^{*}$Definitions of stakeholder roles are provided in Appendix~B.
\end{table*}

\textbf{Phase 2 – Stakeholder-Driven Refinement.}
The initial field inventory derived from the standards analysis was iteratively refined in collaboration with ten stakeholders representing the roles summarized in the last row of Table~\ref{tab:standards_viewpoints_roles}. These roles included Domain Scientists, Data Curators, Semantic Curators, Software Developers, Infrastructure Managers, DT Integrators, Platform Operators, and end Users.
This phase focused on identifying missing integration-relevant attributes, making implicit modeling and deployment assumptions explicit, and resolving ambiguities in field definitions. By incorporating perspectives spanning scientific, technical, and operational domains, the refinement process ensured that the resulting schema reflects practical integration workflows and real-world modeling constraints while preserving the RM-ODP separation of concerns across viewpoints. The refined set of viewpoint-specific metadata fields is summarized in Table~\ref{tab:env_viewpoint_fields}.

\textbf{Phase 3 – Expert Validation.}
Following refinement, the consolidated schema was reviewed by 10 environmental modeling experts representing diverse subfields. Experts assessed the adequacy and completeness of the metadata fields across the five RM-ODP viewpoints and provided structured feedback on coverage and clarity. This validation step ensured that the schema achieved viewpoint-level completeness and domain relevance prior to empirical application. The quantitative and qualitative results of this evaluation are reported in Section~\ref{sec:evaluation}.

\textbf{Phase 4 – Empirical Application.}
The refined schema was subsequently applied to a curated set of environmental DT integration scenarios to assess its adequacy for structured model description and compatibility reasoning. This phase verified that the metadata elements were not only conceptually coherent but also operationally effective for automated mismatch detection. The evaluation results of these applications are reported in Section~\ref{sec:evaluation}.

Through this four-phase process, a domain-specialized set of metadata fields was derived and organized according to the five RM-ODP viewpoints. 

To make the structural instantiation explicit, Table~\ref{tab:env_viewpoint_fields} summarizes the environmental metadata fields grouped by viewpoint. The table lists field names only, emphasizing structural alignment with the generic integration framework. Detailed field definitions and a fine-grained comparison with existing standards are provided in Appendix~A.

\begin{table*}[t]
\centering
\caption{Viewpoint-specific metadata fields for environmental models.}
\label{tab:env_viewpoint_fields}
\footnotesize
\setlength{\tabcolsep}{6pt}
\renewcommand{\arraystretch}{1.25}
\begin{tabular}{p{0.26\linewidth}p{0.68\linewidth}}
\hline
\textbf{RM-ODP Viewpoint} & \textbf{Metadata Fields} \\
\hline

\textbf{Domain (Enterprise)} &
Title; Model Version; Description; Keywords; Model Type; Scope; Purpose; Pattern; Assumptions; Links to Publications and Reports; Conceptual Model Evaluation; Calibration Tools/Data; Validation Capabilities; Sensitivity Analysis; Uncertainty Analysis; Authors' Unique Identifier; Contributor Role \\

\textbf{Information} &
Model's Unique Identifier; Unique Identifier of Submodels; Parameters; Input Datasets; Output; Dimensionality; Spatial Resolution; Variable Spatial Resolution; Time Steps/Temporal Resolution; Variable Temporal Resolution; Resampling/Conversion Policies \\

\textbf{Computational} &
Interface Signature; Error Handling; Integration Pattern \\

\textbf{Engineering} &
Support for Parallel Execution; Execution Constraints; Acknowledgment Protocols; Latency Expectations; Data Synchronization \\

\textbf{Technology} &
Programming Language; Availability of Source Code; Implementation Verification; Software Specification and Requirements; Hardware Specification and Requirements; Execution Instructions; License; Landing Page; Distribution Version \\

\hline
\end{tabular}
\end{table*}

\begin{figure}[h]
\centering
\includegraphics[width=0.7\columnwidth]{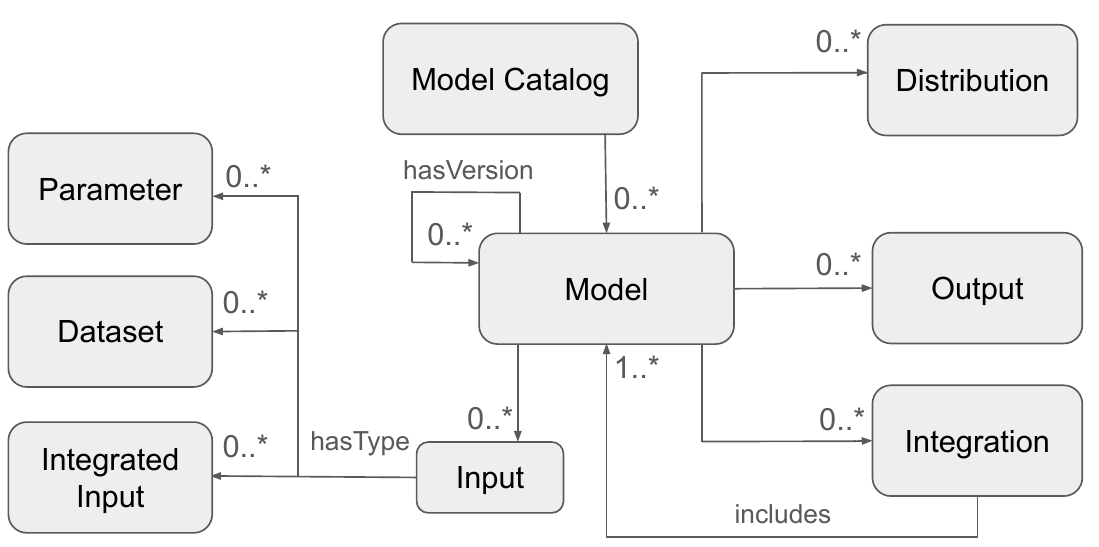}
\caption{Entity-Relationship Diagram of the specialized metadata schema for environmental modeling.}
\label{fig:EDR}
\end{figure}

The ER diagram in Fig.~\ref{fig:EDR} represents the domain-specific realization of the generic integration metamodel introduced in Section~\ref{subsec:metamodel}. Abstract constructs such as \textit{DigitalTwinModel}, \textit{ViewpointDescription}, and \textit{IntegrationSpecification} are instantiated here as concrete environmental metadata entities and relationships. Viewpoint-level abstractions are materialized as structured entities including \textit{Model Catalogue}, \textit{Model Details}, \textit{Input}, \textit{Output}, \textit{Distribution}, \textit{Author}, and \textit{Integration Specification}, while preserving the architectural separation of concerns defined in the generic framework.

The resulting entity structure comprises:
\begin{itemize}
\item \textbf{Model Catalogue} – Manages and indexes model metadata to support discovery, classification, and access control.
\item \textbf{Model Details} – Captures conceptual descriptors such as scope, assumptions, methodology, and spatial and temporal characteristics.
\item \textbf{Distribution} – Describes packaging and deployment details, including dependencies, container images, and execution environments.
\item \textbf{Input} – Specifies required parameters or upstream datasets and models, enabling automated compatibility checks through standardized semantics and units.
\item \textbf{Output} – Records the semantics, structure, and resolution of model outputs, ensuring compatibility with downstream components.
\item \textbf{Author} – Provides contributor identities, affiliations, and provenance information in line with FAIR principles.
\item \textbf{Integration Specification} – Defines integration-specific metadata such as integration pattern, communication mechanisms, and synchronization requirements.
\end{itemize}

This instantiation demonstrates how the generic RM-ODP–based framework can be operationalized for environmental models while preserving its multi-viewpoint structure. By aligning domain-relevant metadata fields with the integration metamodel, the schema provides a structured foundation for automated compatibility analysis and the domain-specific mismatch detection presented in the following subsection.

\subsection{Mismatch Detection Applied to Environmental Modeling}

Building on the generic detection procedure introduced in 
Section~\ref{subsec:mismatch-detection}, we specialize the hybrid 
pattern-aware and LLM-assisted mechanism for the environmental modeling domain 
by binding the abstract constraint categories to the environmental metadata 
fields defined in Section~\ref{subsec:Metamodel-Environmental-Models}. 
In this instantiation, viewpoint-specific metadata elements serve as the 
concrete inputs over which compatibility rules are generated and subsequently 
evaluated by Algorithm~\ref{alg:pattern-aware-detector-merged}. 
The result is a domain-specific taxonomy of rule templates that become 
operational through structured semantic evaluation.

Table~\ref{tab:mismatchs_with_pattern_scope} summarizes the environmental 
mismatch types derived from the generic constraint categories. 
For each mismatch type, the table indicates 
(i) its interpretation in the environmental context, 
(ii) the metadata fields required for rule instantiation, and 
(iii) whether the rule is pattern-agnostic or pattern-specific. 
Each mismatch corresponds to a compatibility rule generated during 
Stage~1 (Pattern-Aware Rule Generation) of 
Algorithm~\ref{alg:pattern-aware-detector-merged}, 
while its final verdict is determined in Stage~2 through LLM-based evaluation.

\textbf{Pattern-Agnostic Mismatches.}
Pattern-agnostic mismatches represent foundational compatibility conditions 
that are instantiated for every integration attempt, regardless of the selected 
integration pattern. These include \textit{Semantic Mismatch} and 
\textit{Conceptual Quality Mismatch}. 
They assess alignment at the Domain viewpoints, ensuring that 
conceptual scope and scientific assumptions are sufficiently consistent before integration planning proceeds. 

\textbf{Pattern-Specific Mismatches.}
Pattern-specific mismatches are activated only when the selected integration 
pattern induces corresponding interaction constraints. 
Consistent with Section~\ref{subsec:mismatch-detection}, 
they are grouped into information-level and runtime-level categories.

\emph{(i) Information-Level Rules.}
Information-level mismatches, including the 
\textit{Information Alignment Mismatch}, are instantiated for the 
data-flow edges activated by the selected integration pattern 
(Algorithm~\ref{alg:pattern-aware-detector-merged}, lines~10–23). 

Let $(A,B)$ denote an ordered pair of models from the available set $M$, 
and let $E_{p}^{AB}$ represent the set of active data-flow edges between 
$A$ and $B$ under pattern $p$. 
An edge $A \rightarrow B$ indicates that model $B$ consumes outputs produced 
by model $A$. The integration pattern determines which exchange directions 
are activated: One-Way patterns evaluate only $A \rightarrow B$ relations; 
Loose and Shared patterns permit bidirectional checks when dependencies exist; 
Integrated and Embedded patterns activate only the directions explicitly 
declared in the integration specification.

For each active exchange relation $(X \rightarrow Y)$, a compatibility rule 
is generated to assess whether the outputs of $X$ can be interpreted as valid 
inputs for $Y$ according to Information-viewpoint metadata. 
These rules evaluate environmental variables, semantics, units, dimensionality, 
spatial and temporal resolution, data-structure descriptors, and provenance 
(see Appendix~\ref{appendix:ODP_metadata_fields}). 

The LLM subsequently evaluates each instantiated rule and produces one of 
three verdicts: \emph{Match}, \emph{Mismatch}, or \emph{Gap}, depending on 
whether the compatibility condition is satisfied or sufficient metadata 
is available for assessment.

\emph{(ii) Runtime-Level Rules.}
Runtime-level mismatches—including 
\textit{Integration Interface Mismatch}, 
\textit{Runtime Coordination Mismatch}, and 
\textit{Technological Mismatch}—capture operational constraints imposed by 
the selected integration pattern. 
These rules assess communication mechanisms, synchronization requirements, 
execution constraints, software dependencies, and technological environments. 
As integration patterns become more tightly coupled, additional runtime rules 
are instantiated and their compatibility conditions become progressively stricter. 
For example, One-Way and Loose patterns primarily require consistent 
data-exchange behavior, whereas Shared and Integrated patterns introduce 
stronger requirements on communication protocols and runtime compatibility. 
Embedded integration further requires that one model can execute within the 
runtime environment of another.

For each instantiated rule—whether pattern-agnostic or pattern-specific—
the LLM evaluates the required compatibility condition based on the available 
environmental metadata. If the necessary metadata elements are missing or 
incomplete, the evaluation yields a \emph{Gap} verdict. If the compatibility 
condition is not satisfied, a \emph{Mismatch} verdict is returned. 
Both outcomes are recorded as structured report entries in the mismatch set 
$\mathcal{B}$, together with explanatory reasoning and potential repair suggestions.

Through this domain-specific binding of abstract constraint categories to 
environmental metadata fields, the generic integration framework becomes 
operational for environmental DT scenarios. The specialization demonstrates 
how systematic, viewpoint-aware rule generation combined with LLM-assisted 
semantic evaluation enables structured diagnosis of semantic, informational, 
and runtime incompatibilities across heterogeneous environmental models, 
while preserving the generality of the underlying integration metamodel.

\begin{table*}[ht]
\centering
\footnotesize
\caption{Integration mismatches grouped by RM-ODP viewpoint and involved metadata fields.}
\label{tab:mismatchs_with_pattern_scope}
\begin{footnotesize}
\begin{tabular}{p{0.25\linewidth}p{0.35\linewidth}p{0.35\linewidth}}
\hline
\textbf{Mismatch Name} & \textbf{Description} & \textbf{Fields Involved} \\
\hline

\multicolumn{3}{l}{\textbf{Enterprise Viewpoint}} \\
\cmidrule(lr){1-1}
Semantic Mismatch &
Misalignment in semantics or purpose. &
Title, Model Version, Description, Keywords, Model Type, Scope, Purpose \& Pattern, Assumptions \\

Conceptual Quality Mismatch &
Model lacks expected scientific or quality evidence. &
Links to Publications \& Reports, Authors’ ID, Conceptual Model Evaluation, Calibration Tools/Data, Validation Capabilities, Sensitivity Analysis, Uncertainty Analysis \\

\hline

\multicolumn{3}{l}{\textbf{Information Viewpoint}} \\
\cmidrule(lr){1-1}
Information Alignment Mismatch &
Check whether exchanged variables match in schema, semantics, and spatial–temporal alignment. &
Information Viewpoint Fields (Table~\ref{tab:env_viewpoint_fields}) \\

\hline

\multicolumn{3}{l}{\textbf{Computational Viewpoint}} \\
\cmidrule(lr){1-1}
Integration Interface Mismatch &
APIs, communication styles, or error-handling behaviours are incompatible, preventing models from interacting through a shared interface. &
Computational Viewpoint Fields (Table~\ref{tab:env_viewpoint_fields}) \\

\hline

\multicolumn{3}{l}{\textbf{Engineering Viewpoint}} \\
\cmidrule(lr){1-1}
Runtime Coordination Mismatch &
Concurrency, ordering, acknowledgment, or latency expectations differ, preventing synchronized execution in distributed environments. &
Engineering Viewpoint Fields (Table~\ref{tab:env_viewpoint_fields}) \\

\hline

\multicolumn{3}{l}{\textbf{Technology Viewpoint}} \\
\cmidrule(lr){1-1}
Technological Mismatch &
The models rely on incompatible implementation platforms or standards, preventing execution within a shared technological environment. &
Technology Viewpoint Fields (Appendix~\ref{appendix:ODP_metadata_fields}) \\

\hline
\end{tabular}
\end{footnotesize}
\end{table*}

\section{Evaluation}
\label{sec:evaluation}

This section evaluates the two core components of the proposed conceptual framework for DT model integration, instantiated in the environmental modeling domain (Section~\ref{sec:case-study-environmental}). While the framework is designed to be domain-independent, environmental modeling provides a demanding validation setting due to heterogeneous model assumptions, diverse data representations, and complex execution environments.

The evaluation addresses:
(1) the RM-ODP-guided metadata schema and its development approach, and
(2) the hybrid pattern-aware and LLM-based mismatch detection mechanism.

The metadata schema is assessed through expert evaluation of field importance and viewpoint-level completeness. The mismatch detector is evaluated empirically on a curated set of published environmental integration configurations. In particular, we explicitly compare a purely rule-based compatibility checking approach with the proposed hybrid pattern-aware mechanism that combines rule-based constraint generation with LLM-assisted reasoning. The evaluation assesses their ability to identify domain-level, information-level, and runtime-level mismatches, as well as specification gaps, prior to implementation.

\subsection{Expert Validation of the RM-ODP–Based Model Metadata Schema}
\label{sec:interview_survey}

To assess the adequacy of the RM-ODP–based metadata schema for DT model integration, a structured survey was conducted with ten experts from diverse subfields of environmental modeling. The objective was to evaluate whether the proposed schema captures the metadata necessary to support systematic integration across viewpoints.

\begin{figure*}[t]
\centering
\includegraphics[width=0.5\columnwidth]{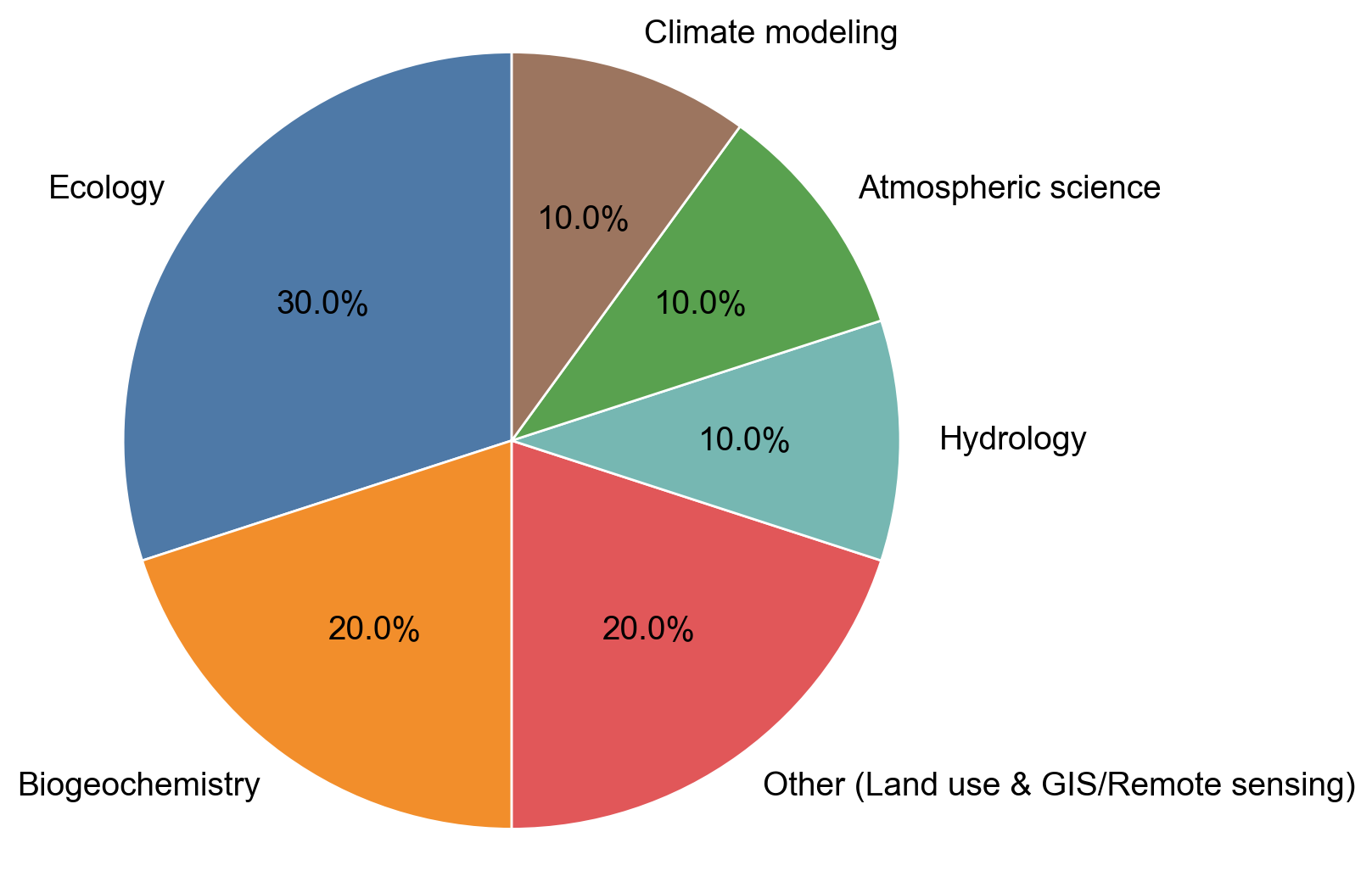}
\caption{Expert participant distribution by disciplinary background in environmental modeling.}
\label{fig:expert_distribution}
\end{figure*}

Participants rated the importance of individual metadata fields for successful integration and evaluated the perceived completeness of the schema within each RM-ODP viewpoint. Fig.~\ref{fig:expert_distribution} presents the disciplinary distribution of participants, ensuring representation across key environmental domains relevant to DT integration.

\begin{figure}[htbp]
\centering
\includegraphics[scale=0.5]{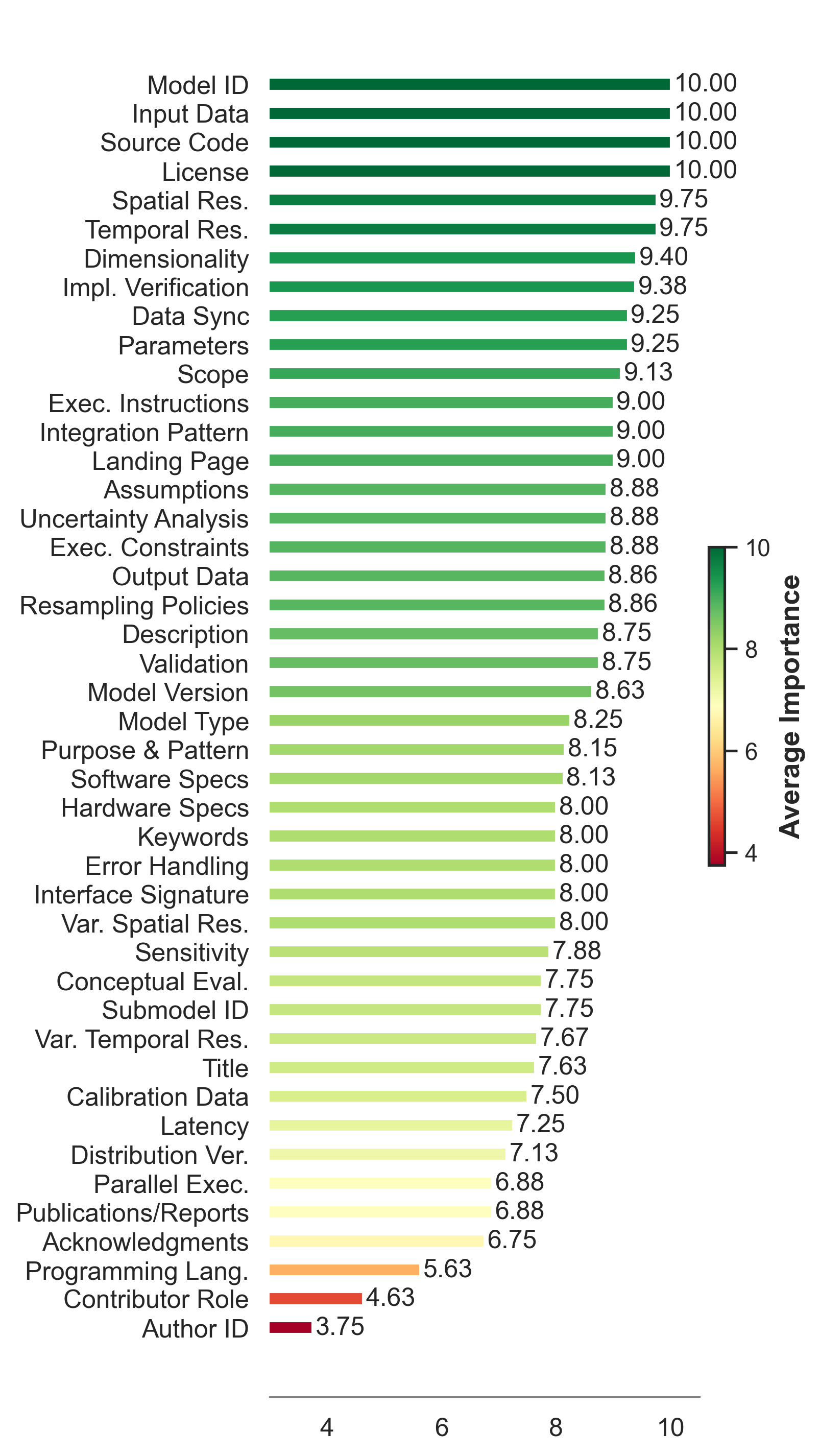}
\caption{Average expert importance ratings for metadata fields relevant to model integration (0 = not important, 10 = essential).}
\label{fig:ModelMetadataResults}
\end{figure}

Fig.~\ref{fig:ModelMetadataResults} shows the average importance ratings for individual metadata fields. The highest-rated elements were \emph{Model’s Unique Identifier}, \emph{Input Datasets}, \emph{Source Code Availability}, and \emph{License}, highlighting the central role of reproducibility, transparency, and traceability in integration scenarios. Structural descriptors such as \emph{Spatial/Temporal Resolution} and \emph{Dimensionality} were also rated highly, underscoring the importance of the \emph{Information} viewpoint for compatibility assessment. In contrast, provenance- and authorship-related fields received comparatively lower ratings, reflecting their relevance for FAIR principles rather than immediate integration feasibility.

\begin{figure*}[h]
\centering
\includegraphics[width=0.5\columnwidth]{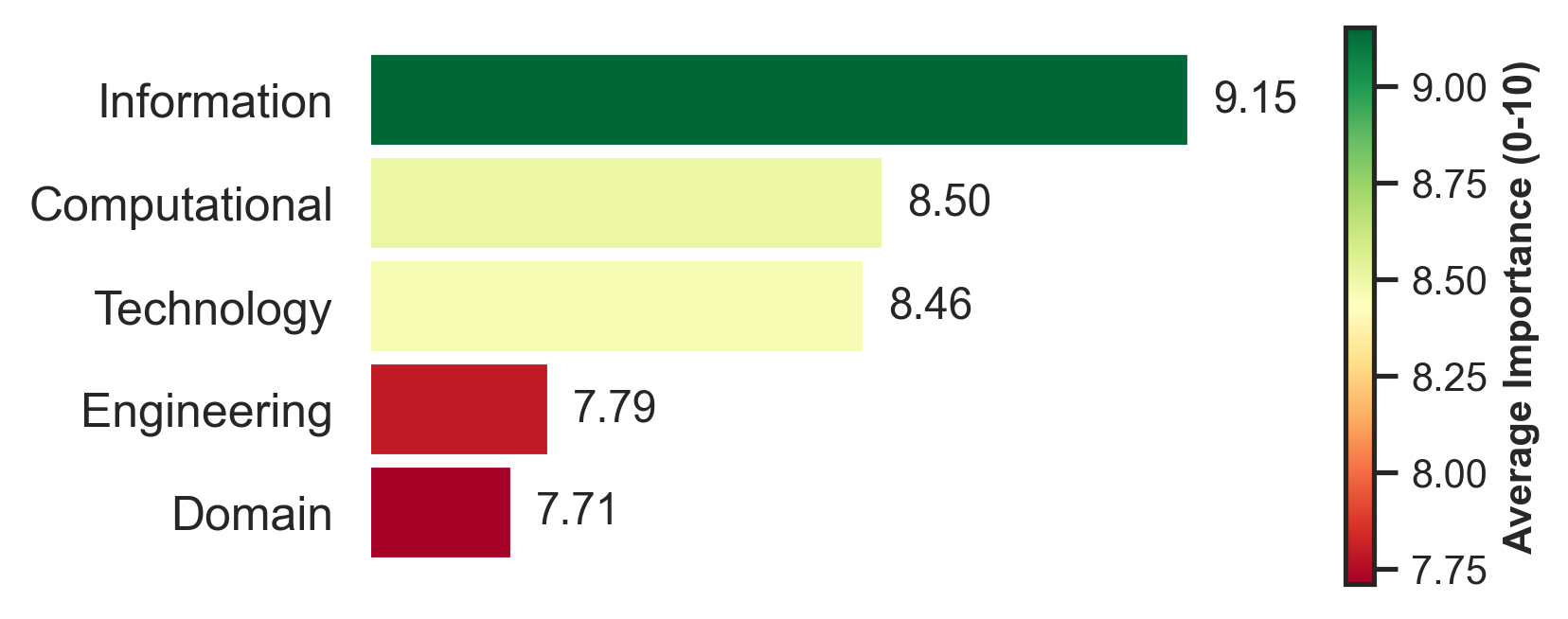}
\caption{Average expert importance of metadata fields by viewpoint (0 = not important, 10 = essential).}
\label{fig:ViewpointsResults}
\end{figure*}

Fig.~\ref{fig:ViewpointsResults} aggregates importance ratings at the viewpoint level. The \emph{Information} viewpoint received the highest overall score, confirming its foundational role in enabling model alignment and reuse. The \emph{Computational} and \emph{Technology} viewpoints followed, reflecting the importance of interface definitions, execution environments, and licensing conditions for practical interoperability. The \emph{Domain} viewpoint received comparatively lower ratings, likely because conceptual assumptions and theoretical scope are often documented in publications rather than formal metadata. The \emph{Engineering} viewpoint was rated lowest, consistent with its stronger relevance during orchestration and deployment rather than early integration design.

\begin{figure}[h]
\centering
\includegraphics[width=0.5\columnwidth]{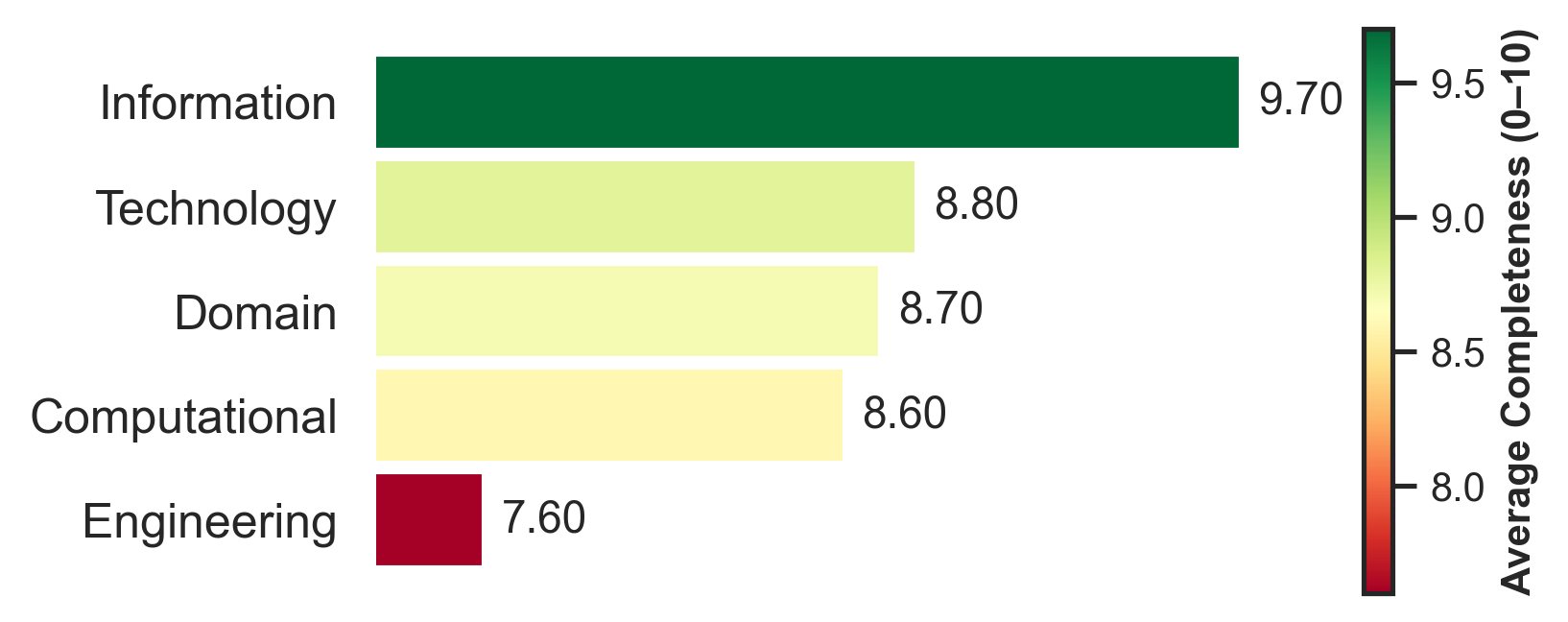}
\caption{Average expert completeness ratings of metadata fields by RM-ODP viewpoint (0 = not complete, 10 = fully complete).}
\label{fig:metamodel_completeness}
\end{figure}

Fig.~\ref{fig:metamodel_completeness} presents completeness ratings per viewpoint. The \emph{Information} viewpoint again achieved the highest scores, indicating broad agreement on the adequacy of variable-level and spatio-temporal descriptors. The \emph{Technology} and \emph{Computational} viewpoints showed solid coverage, while suggesting opportunities for further standardization of execution and interface metadata. The \emph{Domain} viewpoint exhibited disciplinary variation, with ecological and land-use experts assigning higher scores than climate and atmospheric specialists, who emphasized clearer formalization of assumptions and scope. The \emph{Engineering} viewpoint received the lowest completeness ratings, highlighting the need for richer metadata on orchestration, execution coordination, and synchronization to better support automated DT integration.

\subsection{Empirical Validation of the Mismatch Detector}

\subsubsection{Experimental Setup}

For each integration configuration, the RM-ODP–based metadata schema was instantiated at three complementary abstraction levels:

(i) \textbf{Component models (A and B):} 
Structured descriptors of the individual models participating in the integration. Metadata were systematically extracted from the corresponding publications and cross-validated with domain experts to ensure consistency and completeness across the five RM-ODP viewpoints.

(ii) \textbf{Realized integrated model (AB):} 
Metadata describing the composite system as implemented and documented in the original study. This representation reflects the realized integration outcome, including any transformations, adaptations, orchestration mechanisms, or technological alignments introduced during implementation.

(iii) \textbf{Integration Specification (IS):} 
A high-level representation of the intended integration objective, expressed using minimal metadata (e.g., \texttt{title}, \texttt{description}, selected \texttt{integration pattern}). The IS captures the integration intent prior to implementation and serves as the predictive input to the mismatch detection process.

The resulting dataset comprises 44 models organized into 11 integration configurations. Since each realized integrated model combines exactly two component models, each configuration consists of four elements: Model A, Model B, the realized integrated Model AB, and the corresponding IS.

\paragraph{Detection Methods.}
We evaluate two compatibility assessment strategies: (i) a purely rule-based baseline that relies on deterministic constraint checking, and (ii) the proposed hybrid approach that combines rule-based constraint generation with LLM-assisted reasoning.

\textbf{Rule-based baseline.}  
To establish a deterministic baseline, we implemented a simplified rule-based engine that evaluates compatibility by directly comparing the relevant metadata fields of the participating models and the integration specification. Each compatibility rule corresponds to a predefined constraint derived from the RM-ODP viewpoints (e.g., unit consistency, dimensionality alignment, or execution environment compatibility). The rule engine checks these constraints through explicit field comparisons (such as equality checks, range checks, or schema compatibility checks). If the required condition is satisfied, the rule returns \emph{Match}; if it is violated, it returns \emph{Mismatch}. When the required metadata is missing or insufficient to evaluate the rule, the outcome is classified as \emph{Gap}. This baseline represents a purely deterministic compatibility checker without contextual reasoning.

\textbf{LLM-assisted reasoning models.}  
To evaluate the effect of contextual reasoning, we tested several instruction-tuned large language models representing different model families and parameter scales: (i) OpenAI GPT-OSS-120B, (ii) Mistral Small 3.2 24B Instruct, and (iii) Llama 3.3 70B Instruct (AWQ). These models were selected because they are openly deployable instruction-tuned models suitable for structured reasoning tasks, represent different parameter scales (24B–120B), and were available through the same inference infrastructure, ensuring consistent experimental conditions.

For each compatibility rule generated by the pattern-aware rule generator, the LLM receives the rule description together with the relevant metadata context (models A, B, and the IS). The model then produces a structured evaluation consisting of a verdict (\emph{Match}, \emph{Mismatch}, or \emph{Gap}) together with a short explanation.

\paragraph{Evaluation Procedure.}

We evaluate the detector in a predictive setting by comparing its outputs on the IS with the corresponding realized integrated model (AB). For each configuration, both the static rule-based baseline and the LLM-assisted variants assess compatibility between the component models (A, B) and the IS, producing predicted outcomes (\emph{Match}, \emph{Mismatch}, or \emph{Gap}) that represent anticipated integration issues prior to implementation.

These predictions are then compared with the metadata of the realized integrated model (AB), which serves as the ground truth for integration decisions. Based on this comparison, each case is classified as a true detection, false detection, missed detection, or correct match. This setup enables a quantitative evaluation of detector performance across viewpoints and integration patterns.

\subsubsection{Results and Analysis}
This comparison directly evaluates the added value of LLM-assisted reasoning beyond rule-based compatibility checking in the proposed framework. We evaluate the predictive performance of the mismatch detector by comparing the predicted compatibility outcomes with the realized integration decisions observed in the integrated models (AB).

Table~\ref{tab:detection_performance} summarizes the aggregate detection performance across all configurations. The rule-based baseline achieves relatively high recall but very low precision, indicating that although many potential conflicts are detected, a substantial number of false positives are produced. This behavior reflects the conservative nature of purely rule-based checks, which lack contextual interpretation.

In contrast, LLM-assisted variants significantly improve the balance between precision and recall. GPT-OSS-120B achieves the strongest overall performance (F1 = 0.86), followed by Mistral Small 3.2 24B Instruct (F1 = 0.61) and Llama 3.3 70B Instruct (AWQ) (F1 = 0.50). These results demonstrate that incorporating contextual reasoning enables more accurate compatibility assessment compared with purely static rule evaluation.

\begin{table}[h]
\centering
\caption{Detection performance comparison between the rule-based baseline and the LLM-assisted detector.}
\label{tab:detection_performance}
\footnotesize
\begin{tabular}{lcccc}
\hline
\textbf{Method} & \textbf{Accuracy} & \textbf{Precision} & \textbf{Recall} & \textbf{F1-score} \\
\hline
Rule-based (Static) & 0.80 & 0.22 & 0.86 & 0.35 \\
OpenAI GPT-OSS-120B & 0.98 & 0.86 & 0.86 & 0.86 \\
Mistral Small 3.2 24B Instruct & 0.94 & 0.50 & 0.77 & 0.61 \\
Llama 3.3 70B Instruct (AWQ) & 0.90 & 0.37 & 0.77 & 0.50 \\
\hline
\end{tabular}
\end{table}

\paragraph{Class-Conditional Prediction Behavior.}

\begin{figure*}[h]
\centering
\includegraphics[width=\textwidth]{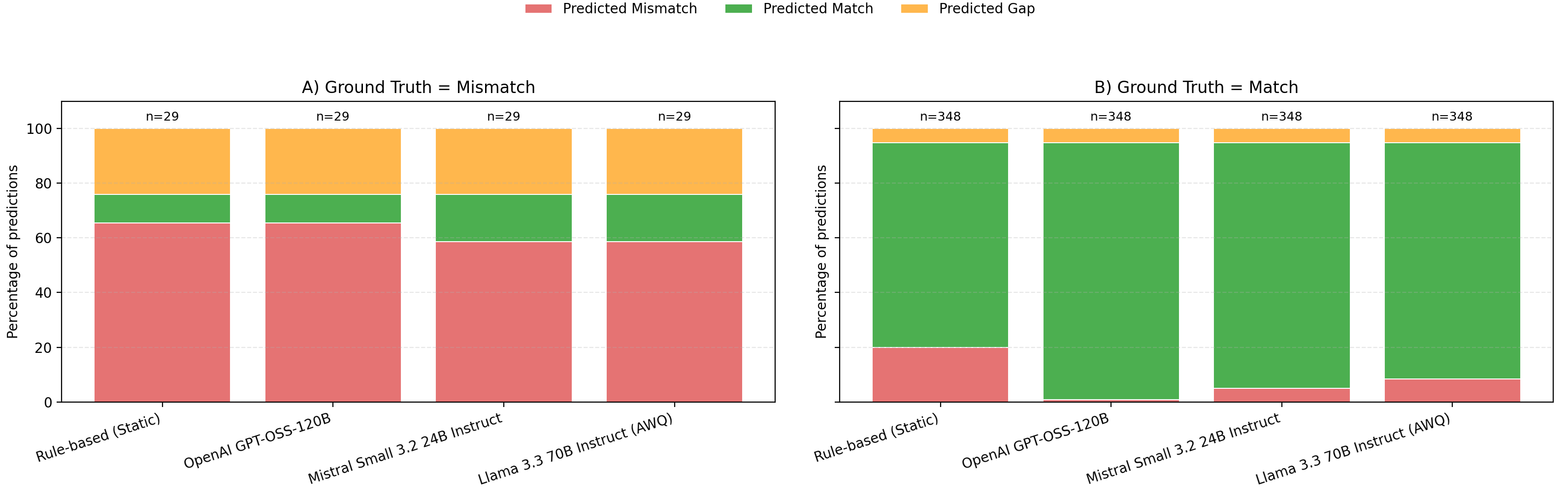}
\caption{Class-conditional prediction distribution across methods. Left: cases where the ground truth indicates a mismatch. Right: cases where the ground truth indicates a match.}
\label{fig:gt_match_vs_mismatch}
\end{figure*}

To further understand detector behavior, we analyze the distribution of predicted outcomes across ground-truth classes. Fig.~\ref{fig:gt_match_vs_mismatch} presents the class-conditional prediction distribution by separating cases where the ground truth corresponds to a \emph{Mismatch} from those where the ground truth corresponds to a \emph{Match}.

When the ground truth indicates a mismatch ($n=29$), most incompatibilities are correctly detected by all methods. 
Only a very small number of cases are incorrectly predicted as \emph{Match}, indicating that the detector rarely overlooks real conflicts when sufficient metadata is available and no specification gaps are present.

When the ground truth indicates a match ($n=348$), the advantage of the LLM-based methods over the rule-based method becomes clear. The LLM-based methods mostly predict \emph{Match}, indicating low false-positive rates. Among them, GPT-OSS-120B shows the most stable behavior, producing the fewest incorrect mismatch predictions. This suggests that LLM-assisted reasoning can detect conflicts accurately while still recognizing valid integrations reliably.

\paragraph{Performance Across RM-ODP Viewpoints.}

\begin{figure*}[h]
\centering
\includegraphics[width=0.85\textwidth]{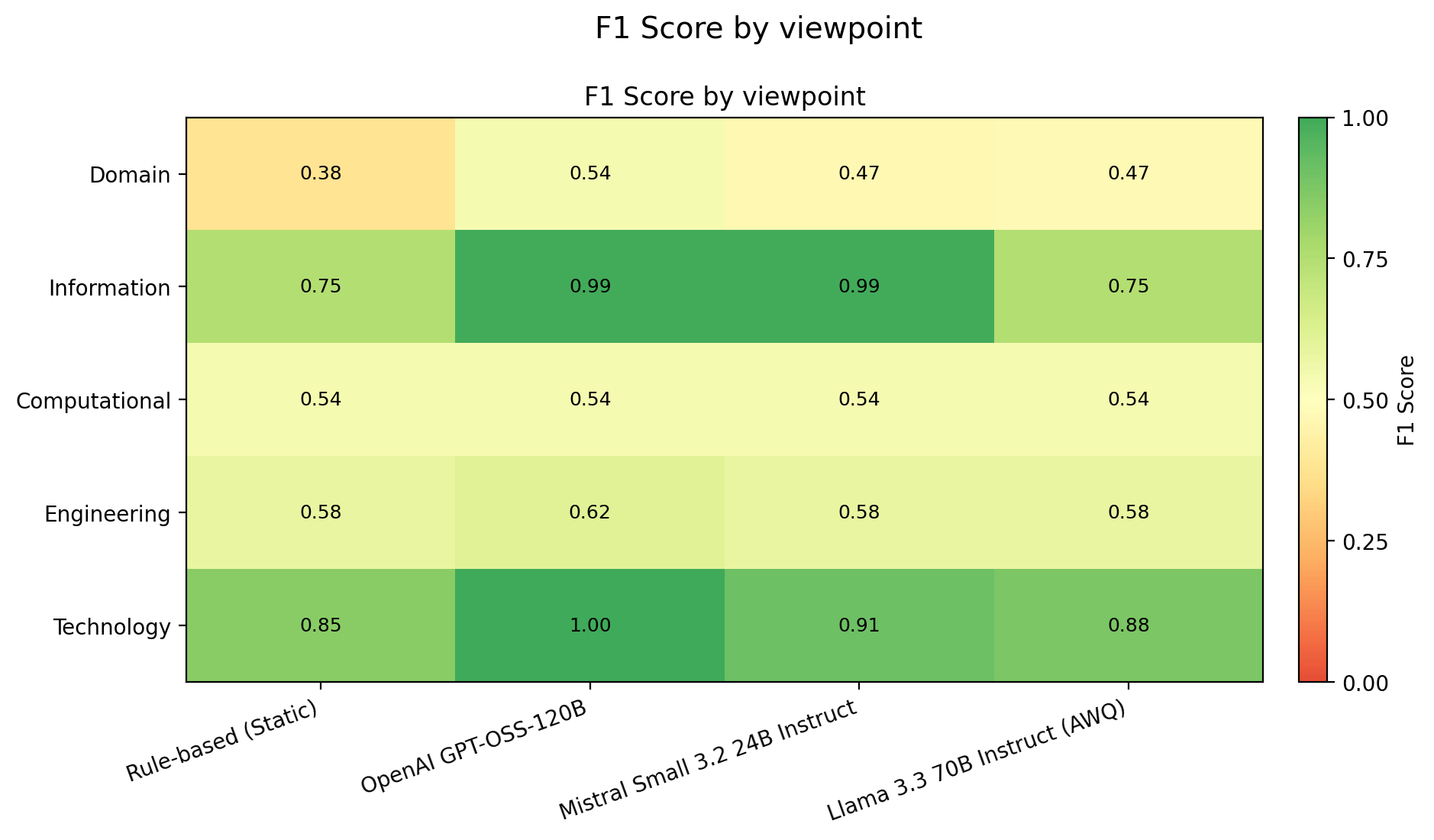}
\caption{F1 score across RM-ODP viewpoints for the rule-based baseline and LLM-assisted detectors.}
\label{fig:viewpoint_macroF1}
\end{figure*}

To analyze the difficulty of compatibility reasoning across viewpoints, Fig.~\ref{fig:viewpoint_macroF1} presents F1 scores disaggregated by RM-ODP viewpoint.

 The results reveal systematic differences in viewpoint difficulty, as well as a clear advantage of the hybrid LLM-assisted approach over purely rule-based evaluation.  The \textbf{Technology} viewpoint achieves the highest scores across all methods (reaching 1.00 for GPT-OSS-120B), reflecting that concrete implementation metadata—such as programming languages, software environments, and versioning—can be evaluated through explicit and deterministic comparisons, where both rule-based and LLM-assisted methods perform reliably. The \textbf{Information} viewpoint also exhibits strong performance, with LLM-assisted methods achieving near-perfect scores (up to 0.99). This indicates that combining structured rules with contextual reasoning enables accurate assessment of data schema compatibility, unit alignment, and structural relationships.

In contrast, the \textbf{Computational} and \textbf{Engineering} viewpoints show moderate performance (around 0.54–0.62), suggesting that execution coordination, interface behavior, and runtime orchestration are more difficult to infer from metadata alone and often require additional contextual interpretation. The \textbf{Domain} viewpoint yields the lowest scores (0.38–0.54), highlighting the inherent ambiguity of conceptual model descriptions and scientific assumptions. 

Notably, LLM-assisted methods consistently outperform the rule-based baseline in both the \textbf{Domain} and \textbf{Information} viewpoints. This improvement can be attributed to the ability of LLMs to interpret unstructured and semi-structured metadata, capture implicit relationships, and reason over contextual descriptions that cannot be fully expressed through deterministic rules alone.

\paragraph{Performance Across Integration Patterns.}

\begin{figure*}[h]
\centering
\includegraphics[width=0.85\textwidth]{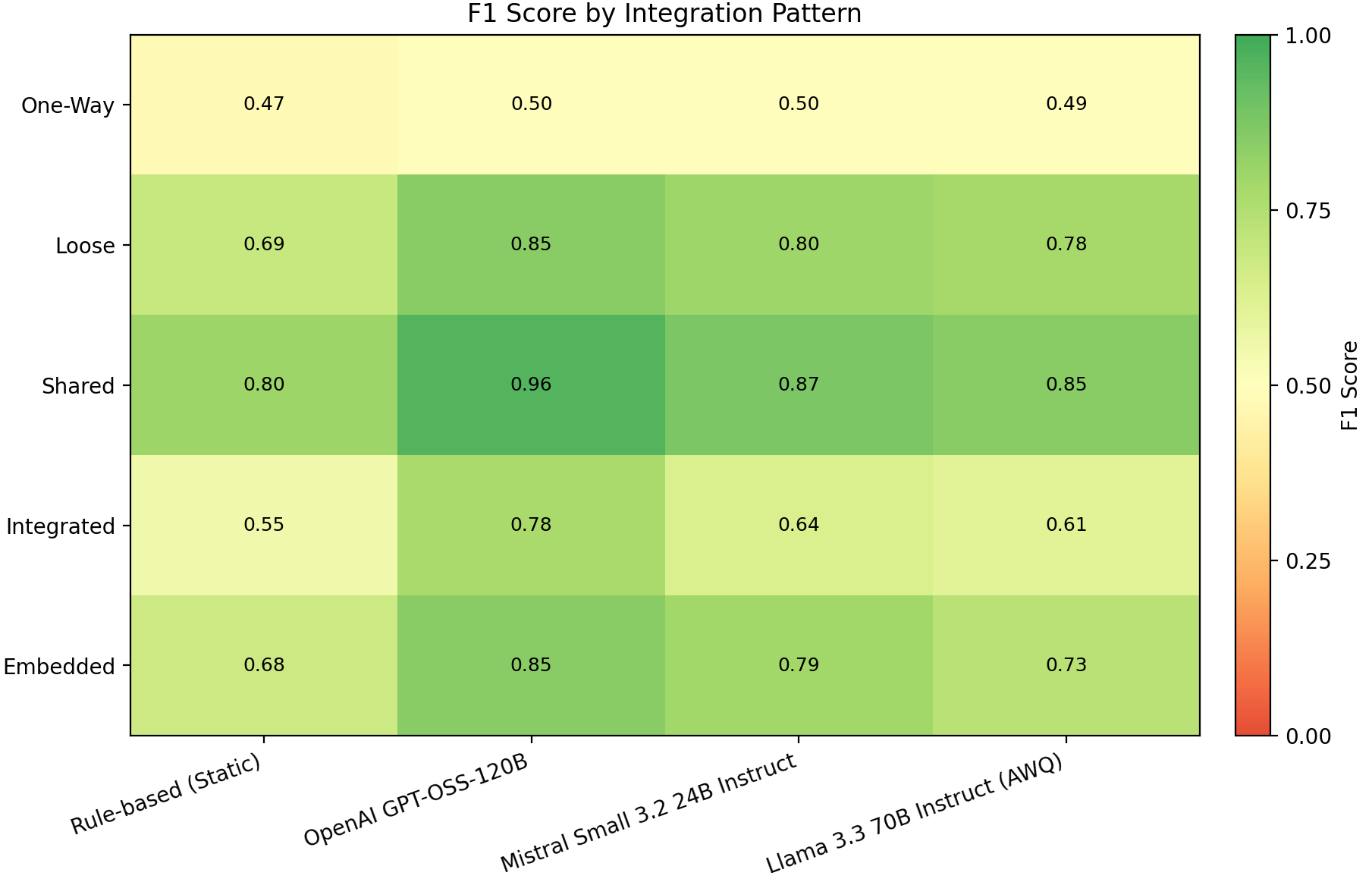}
\caption{F1 score across integration patterns for the rule-based baseline and LLM-assisted detectors.}
\label{fig:pattern_macroF1}
\end{figure*}

To better understand structural influences on detector performance, Fig.~\ref{fig:pattern_macroF1} reports F1-scores disaggregated by integration pattern. The results show that compatibility prediction performance is not solely determined by structural complexity, but also by the availability and explicitness of integration-relevant metadata across patterns.

The results reveal clear structural differences. The \textbf{Shared} pattern achieves the highest performance across all methods (up to 0.96 for GPT-OSS-120B), suggesting that compatibility constraints are easier to reason about when models operate over shared data representations.

The \textbf{Loose} pattern also demonstrates strong performance, particularly for GPT-OSS-120B (0.85), reflecting the relative predictability of loosely coupled orchestration mechanisms.

More tightly coupled patterns such as \textbf{Integrated} and \textbf{Embedded} show moderately lower performance, indicating that deeper runtime coordination and implementation dependencies increase the difficulty of accurate pre-implementation reasoning.

Finally, the \textbf{One-Way} pattern consistently yields the lowest F1 values (approximately 0.47–0.50 across methods). Although structurally simpler, these integrations are typically described with limited specification detail, leaving many semantic and representational assumptions implicit. As a result, they exhibit more frequent metadata gaps, reducing the observable evidence available for compatibility reasoning and making it more difficult to accurately detect inconsistencies. This indicates that the lower performance is primarily driven by metadata incompleteness rather than limitations of the mismatch detection mechanism itself. 

This finding further indicates that improving metadata completeness—potentially through automated metadata harvesting and extraction techniques, as discussed in Section~\ref{sec:discussion}—could enhance reasoning performance by providing richer contextual information for compatibility assessment.

Across all integration patterns, GPT-OSS-120B consistently achieves the strongest performance, indicating more robust reasoning under heterogeneous integration constraints.

\paragraph{Distribution of Detected Mismatch Types.}

\begin{figure*}[h]
\centering
\includegraphics[width=0.7\textwidth]{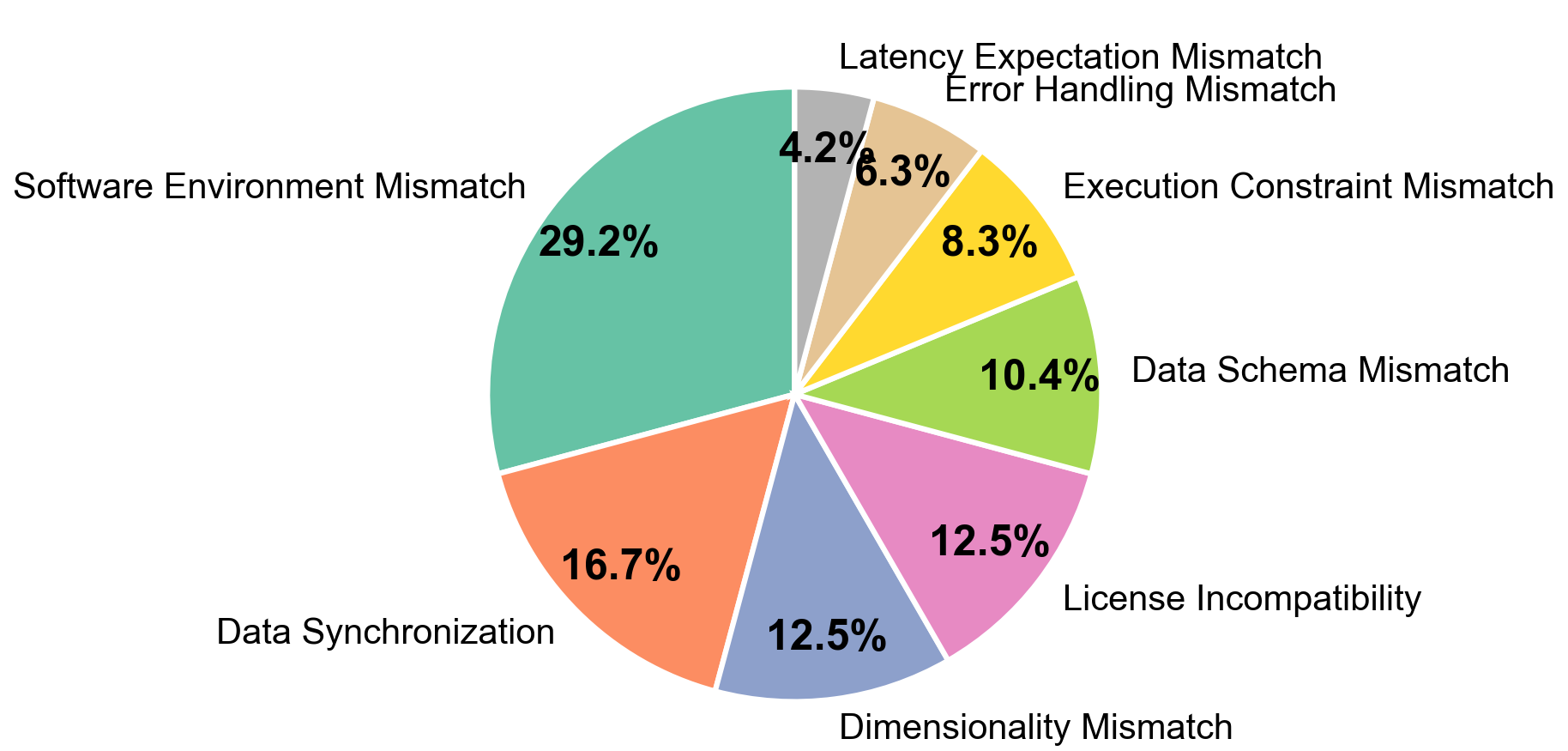}
\caption{Distribution of mismatch types across all evaluated integration configurations.}
\label{fig:bottleneck_pie}
\end{figure*}

Finally, Fig.~\ref{fig:bottleneck_pie} summarizes the distribution of mismatch categories detected by the best-performing model (GPT-OSS-120B).

Software environment mismatches represent the largest category (29.2\%), reflecting differences in runtime dependencies and deployment environments across independently developed models. Data synchronization issues (16.7\%) and dimensionality mismatches (12.5\%) are also common, indicating challenges in aligning temporal update frequencies and spatial representations.

Other categories include licensing incompatibilities (12.5\%), schema mismatches (10.4\%), execution constraints (8.3\%), and error-handling inconsistencies (6.3\%). Latency expectation mismatches appear less frequently (4.2\%), but remain relevant in tightly coupled integrations requiring near-real-time coordination.

\paragraph{Overall Interpretation.}

Taken together, these results demonstrate that combining structured rule evaluation with LLM-assisted reasoning significantly improves predictive alignment with realized integration outcomes. The improvements are particularly evident in semantically rich viewpoints (Domain and Information) and moderately coupled integration patterns (Shared and Loose), while tightly coupled runtime configurations remain comparatively more challenging.

Taken together, these results demonstrate that combining structured rule evaluation with LLM-assisted reasoning significantly improves predictive alignment with realized integration outcomes compared with purely rule-based approaches. The improvement is most pronounced in semantically rich and under-specified settings—particularly in the Domain and Information viewpoints—where compatibility depends on interpreting implicit assumptions and heterogeneous metadata. In contrast, purely rule-based methods remain effective for explicit, well-structured constraints but suffer from high false-positive rates due to lack of contextual interpretation. These findings confirm that the hybrid approach provides a more robust and reliable mechanism for early integration feasibility assessment in heterogeneous DT ecosystems.

\section{Discussion}
\label{sec:discussion}

The evaluation results demonstrate that the proposed multi-viewpoint modeling framework provides a structured basis for reasoning about model integration at the modeling level rather than solely at the execution level. By organizing integration-relevant metadata across RM-ODP viewpoints, the framework makes cross-view assumptions explicit and enables systematic, pre-implementation compatibility assessment. This is particularly important given the inherent difficulty of integration in heterogeneous DT systems, where dependencies span multiple abstraction layers and are often only revealed during implementation. In this context, the modeling-centric perspective introduced in this work constitutes its primary conceptual contribution.

The environmental modeling case study confirms the feasibility of the proposed framework and highlights broader implications that extend beyond the evaluated domain, as discussed below.

\subsection{Metadata Population and Automation Challenges}

Populating the proposed model metadata schema for existing environmental models proved to be one of the most time-consuming aspects of the evaluation. Many required descriptors—such as assumptions, input/output structures, execution dependencies, and synchronization constraints—were not available in structured form and had to be reconstructed from publications, supplementary materials, or source code. Even when documentation existed, extracting a complete and consistent set of descriptors was often infeasible due to ambiguities or missing information. The substantial number of metadata gaps observed in the empirical results confirms that manual collection alone cannot ensure completeness or consistency across heterogeneous models.

These findings highlight the need for automated \emph{metadata harvesters} capable of extracting structured descriptors directly from model artifacts, documentation, and simulation workflows. While such tools are common in the data domain—for example, flexible metadata harvesters for dataset catalogs~\cite{lu2025flexible}—an equivalent capability for computational models remains largely absent. Techniques such as static code analysis, annotation mining, workflow introspection, and natural language processing could significantly reduce manual effort and improve metadata coverage. Emerging Large Language Models (LLMs) offer additional potential by interpreting unstructured documentation to infer dependencies, parameters, assumptions, and data flows. Incorporating LLM-assisted extraction mechanisms would enable scalable and adaptive metadata generation for large model ecosystems, extending the framework’s applicability beyond environmental modeling to broader scientific and industrial DT contexts.

\subsection{Limits and Opportunities of LLM-Assisted Compatibility Reasoning}

The empirical evaluation (Section~\ref{sec:evaluation}) shows that incorporating LLM-assisted reasoning improves compatibility detection compared with purely rule-based evaluation. The improvements are particularly visible in viewpoints that require contextual interpretation, such as the \emph{Domain} and \emph{Information} viewpoints.

Rule-based checks remain effective for structural constraints that can be expressed through explicit metadata comparisons, such as unit consistency, dimensionality alignment, or software environment compatibility. However, purely deterministic rules often struggle to capture conceptual relationships between models. Domain-level compatibility frequently depends on interpreting model objectives, assumptions, and methodological scope, which are typically described in narrative form rather than as structured attributes.

LLM-assisted reasoning helps address this limitation by interpreting textual descriptions together with the surrounding metadata context. By combining rule descriptions with viewpoint-structured metadata for the participating models and the integration specification, the LLM can reason about compatibility conditions that cannot easily be expressed through deterministic constraints alone.

At the same time, the evaluation highlights an important limitation of LLM-based reasoning: its effectiveness depends strongly on the availability and quality of contextual information. When key metadata descriptors are missing or incomplete, the LLM may produce uncertain outcomes or return \emph{Gap} predictions. In such cases, the model cannot infer integration feasibility reliably without sufficient contextual evidence.

\subsection{From Detection to Resolution}

The empirical analysis also reveals that many mismatches detected by the framework correspond to integration challenges that were ultimately resolved during implementation in the analyzed studies. In practice, such mismatches are often addressed through data transformations, interface adapters, or orchestration mechanisms.

This observation suggests that compatibility assessment should not only detect inconsistencies but also support reasoning about feasible resolution strategies. Within the proposed framework, such strategies can be conceptualized as explicit mediation constructs introduced at the modeling level. Examples include data harmonization pipelines, intermediate abstraction layers, interface adapters, and orchestration components that reconcile differences between models.

The LLM-assisted reasoning component further contributes to this process by generating candidate suggestions for resolving detected mismatches based on the surrounding metadata context. Rather than merely reporting incompatibilities, the framework can therefore provide actionable guidance for integration design, suggesting potential transformations or architectural adaptations that enable successful model coupling.

\subsection{Cross-View Quality and Structural Implications}

The viewpoint-based structuring reveals that integration quality is inherently multi-dimensional. Semantic validity (Domain viewpoint), structural consistency (Information viewpoint), behavioral coordination (Engineering viewpoint), and technological compatibility (Technology viewpoint) represent distinct yet interdependent quality dimensions.

The empirical results further indicate that tightly coupled integration patterns are associated with a higher concentration of detected mismatches and metadata gaps. This suggests that coupling strength influences the complexity of cross-view alignment. By organizing constraints across viewpoints, the framework enables traceable reasoning about how inconsistencies propagate across abstraction layers. For example, ambiguities at the Domain viewpoint may manifest as structural inconsistencies at the Information viewpoint or runtime coordination issues at the Engineering viewpoint.

This structural traceability constitutes an advantage over execution-centric approaches, which typically diagnose failures reactively rather than enabling modeling-level reasoning prior to deployment.

\subsection{Generalizability and Threats to Validity}

The evaluation was conducted within the environmental modeling domain, a heterogeneous and integration-intensive setting. Although the metamodel itself is domain-agnostic and structured around architectural viewpoints rather than domain semantics, additional empirical studies across other domains—such as manufacturing, cyber-physical systems, or healthcare—are necessary to assess schema extensibility and modeling adequacy.

A further limitation concerns metadata completeness. The effectiveness of mismatch detection depends on the availability and quality of viewpoint-aligned descriptors. Incomplete or inconsistently documented metadata may reduce detection precision, as observed particularly at the Domain viewpoint.

Despite these limitations, the viewpoint-based structuring and pattern-aware detection logic remain independent of domain-specific semantics, suggesting that the framework can be systematically instantiated in other integration contexts.
\section{Conclusion}  
\label{sec:conclusion}

This paper introduces a multi-viewpoint integration modeling framework for heterogeneous DT ecosystems, grounded in the RM-ODP architectural viewpoints. The proposed \textit{Model Metadata Schema} structures integration-relevant knowledge across domain, information, computational, engineering, and technology perspectives, enabling integration assumptions to be represented explicitly and in a machine-interpretable form. On top of this modeling layer, the \textit{Mismatch Detector} operationalizes viewpoint-aware and pattern-aware compatibility reasoning through a hybrid mechanism that combines deterministic rule generation with structured LLM-assisted evaluation of compatibility constraints.

Through expert validation and empirical evaluation in the environmental modeling domain, the feasibility and practical applicability of the framework were demonstrated. The case study illustrates how elevating integration concerns to the modeling layer enables structured compatibility reasoning beyond execution-centric interoperability approaches. The empirical results further indicate that incorporating LLM-assisted reasoning improves predictive alignment with realized integration outcomes compared with purely rule-based compatibility checks.

By making cross-view dependencies explicit, the framework provides a principled basis for systematic model discovery, feasibility assessment, and integration analysis in heterogeneous DT ecosystems.

Future work will focus on extending the modeling abstraction to incorporate formalized mediation constructs for mismatch resolution, allowing adaptation strategies to be represented and reasoned about within the metamodel itself. Additional research will investigate automated support for metadata generation and semantic alignment, including ontology-based reasoning and language-model–assisted interpretation, while preserving analyzability and traceability across viewpoints. These directions aim to further strengthen the methodological foundations of model integration in DT ecosystems and support scalable, transparent, and reproducible integration practices.

\bmhead{Acknowledgements}
This research was primarily supported by the Dutch NWO LTER-LIFE program and was made possible in part through funding from several European Union projects, including EVERSE (101129744), ENVRI-Hub Next (101131141), BlueCloud-2026 (101094227), OSCARS (101129751), and the LifeWatch ERIC project.


\bmhead{Supplementary information}
`Not applicable'



\section*{Declarations}
`Not applicable'








\begin{appendices}

\section{RM-ODP Metadata Fields for Environmental Models}
\label{appendix:ODP_metadata_fields}
\setcounter{table}{0}
\renewcommand{\thetable}{A\arabic{table}}
Table~\ref{tab:ODP_metadata_fields} provides the detailed metadata field definitions
associated with each RM-ODP viewpoint and, as referenced in
Section~\ref{sec:case-study-environmental}, also compares the presence of these
fields across relevant domain-specific standards.

  \footnotesize  
  \setlength{\tabcolsep}{4pt}
  \renewcommand{\arraystretch}{1.05}

  \begin{longtable}{p{3cm}p{5cm}cccc}
\caption{Metadata fields grouped by RM-ODP viewpoint. 
Symbols: ``–'' = no coverage, ``(\checkmark)'' = partial coverage, ``\checkmark'' = full coverage.}
\label{tab:ODP_metadata_fields} \\
  \label{tab:ODP_metadata_fields} \\
  \toprule
  \textbf{Field} & \textbf{Description} & \textbf{ODD} & \textbf{TRACE} & \textbf{SED-ML} & \textbf{SBML L3} \\
  \midrule
  \endfirsthead
  \multicolumn{6}{c}{{\tablename\ \thetable{} -- continued from previous page}}\\
  \toprule
  \textbf{Field} & \textbf{Description} & \textbf{ODD} & \textbf{TRACE} & \textbf{SED-ML} & \textbf{SBML L3} \\
  \midrule
  \endhead
  \midrule
  \multicolumn{6}{r}{{Continued on next page}}\\
  \endfoot
  \bottomrule
  \endlastfoot

\multicolumn{2}{@{}l}{\textbf{Domain (Enterprise) viewpoint}} & & & & \\ \cmidrule(lr){1-1}
\texttt{Title} & The formal title of the model, simulation, or component. & (\checkmark) & - & (\checkmark) & \checkmark \\
\texttt{Model Version} & Version of the model’s scientific design. & (\checkmark) & (\checkmark) & (\checkmark) & (\checkmark) \\
\texttt{Description} & Summary of the model's structure and application. & \checkmark & \checkmark & (\checkmark) & \checkmark \\
\texttt{Keywords} & Controlled vocabulary terms or keywords describing domain, phenomena, or process type. & - & - & - & - \\
\texttt{Model Type} & Scientific paradigm or computational approach (e.g., empirical, mechanistic, hybrid, data-driven). & (\checkmark) & - & - & (\checkmark) \\
\texttt{Scope} & Spatial and temporal scales of applicability, including domain coverage (e.g., global, catchment, basin, grid cell). & (\checkmark) & (\checkmark) & - & - \\
\texttt{Purpose \& Pattern} & Scientific or operational purpose and high-level process logic (e.g., forecasting, hindcasting, management scenario, policy analysis). & \checkmark & \checkmark & - & - \\
\texttt{Assumptions} & Conceptual or theoretical assumptions. & (\checkmark) & \checkmark & - & (\checkmark) \\
\texttt{Links to Publications and Reports} & Supporting literature references. & (\checkmark) & (\checkmark) & \checkmark & \checkmark \\
\texttt{Conceptual Model Evaluation} & Justification and theoretical basis. & (\checkmark) & \checkmark & - & (\checkmark) \\
\texttt{Calibration Tools/Data} & Methods, algorithms, and datasets used for calibration or data assimilation. & (\checkmark) & \checkmark & (\checkmark) & - \\
\texttt{Validation Capabilities} & Documented validation data, tests, and readiness for independent scientific validation. & (\checkmark) & \checkmark & (\checkmark) & - \\
\texttt{Sensitivity Analysis} & Analyses quantifying how outputs respond to parameter or input variation. & - & \checkmark & (\checkmark) & - \\
\texttt{Uncertainty Analysis} & Methods and metrics for uncertainty propagation or quantification in outputs. & - & \checkmark & (\checkmark) & - \\
\texttt{Authors' Unique Identifier} & Contributor IDs (e.g., ORCID). & - & - & (\checkmark) & \checkmark \\
\texttt{Contributor Role} & Function of each contributor in the scientific work. & - & (\checkmark) & (\checkmark) & (\checkmark) \\
\midrule

\multicolumn{2}{@{}l}{\textbf{Information viewpoint}} & & & & \\ \cmidrule(lr){1-1}
\texttt{Model's Unique Identifier} & A persistent ID for the model (DOI, UUID, etc.). & - & - & \checkmark & \checkmark \\
\texttt{Unique Identifier of Submodels} & ID for subcomponents within a composite model. & (\checkmark) & - & \checkmark & (\checkmark) \\ 
\texttt{Parameters} & Structured metadata for model parameters including names, semantics, units, valid ranges, and calibration status. & \checkmark & (\checkmark) & \checkmark & \checkmark \\
\texttt{Input Datasets} & External data dependencies (e.g., forcing, boundary conditions, observations), including data type, temporal cadence, and source. & (\checkmark) & \checkmark & - & - \\
\texttt{Output} & Variables or fields produced by the model, with defined semantics, units, and temporal/spatial characteristics. & (\checkmark) & (\checkmark) & \checkmark & (\checkmark) \\
\texttt{Dimensionality} & Spatial and temporal dimensionality (e.g., 0D–3D, time-stepped or event-driven). & \checkmark & - & - & - \\
\texttt{Spatial Resolution} & Native spatial resolution. & \checkmark & - & - & - \\
\texttt{Variable Spatial Resolution} & Indicates whether the model supports dynamic spatial resolution. & (\checkmark) & - & - & - \\
\texttt{Time Steps/Temporal Resolution} & Native temporal resolution of updates. & \checkmark & (\checkmark) & \checkmark & - \\
\texttt{Variable Temporal Resolution} & Indicates if the model supports adaptive time steps. & (\checkmark) & - & - & - \\
\texttt{Resampling/Conversion Policies (per variable)} & Permitted transformations for interoperability (e.g., unit conversion, coordinate/projection changes, time conversions). & - & - & - & - \\
\midrule

\multicolumn{2}{@{}l}{\textbf{Computational viewpoint}} & & & & \\ \cmidrule(lr){1-1}
\texttt{Interface Signature} & Formal specification of callable operations exposed by the model component, including operation names, required input arguments, and returned outputs. & - & - & - & - \\
\texttt{Error Handling} & Procedures for managing exceptions, numerical instability, or failed data exchanges. & - & (\checkmark) & - & - \\
\texttt{Integration Pattern} & Type of model linkage. & - & - & - & - \\
\\
\midrule

\multicolumn{2}{@{}l}{\textbf{Engineering viewpoint}} & & & & \\ \cmidrule(lr){1-1}
\texttt{Support for Parallel Execution} & Logical and infrastructural support for concurrent execution. & - & - & - & - \\
\texttt{Execution Constraints} & Constraints for execution order/timing. & - & (\checkmark) & - & (\checkmark) \\
\texttt{Acknowledgment Protocols} & Mechanisms confirming message delivery or data transfer between integrated components. & - & - & - & - \\
\texttt{Latency Expectations} & Expected communication latency or acceptable delay between component interactions. & - & - & - & - \\
\texttt{Data Synchronization} & Strategies for synchronizing data across models (synchronous, asynchronous, or event-driven). & - & - & - & - \\
\midrule

\multicolumn{2}{@{}l}{\textbf{Technology viewpoint}} & & & & \\ \cmidrule(lr){1-1}
\texttt{Programming Language} & Implementation language (e.g., Python, R). & - & - & (\checkmark) & - \\
\texttt{Availability of Source Code} & Access to the actual codebase. & (\checkmark) & - & (\checkmark) & - \\
\texttt{Implementation Verification} & Tests confirming the correctness of the implemented code. & - & \checkmark & (\checkmark) & - \\
\texttt{Software Specification and Requirements} & Required software environment/tools. & - & (\checkmark) & (\checkmark) & - \\
\texttt{Hardware Specification and Requirements} & Minimum/optimal hardware resources (CPU, memory, GPU, etc.). & - & - & - & - \\
\texttt{Execution Instructions} & Steps to run the implemented model. & - & (\checkmark) & \checkmark & - \\
\texttt{License} & Legal reuse terms. & (\checkmark) & - & - & (\checkmark) \\
\texttt{Landing Page} & Web-accessible main page for the model/software. & - & - & - & \checkmark \\
\texttt{Distribution Version} & Version of the distributed implementation package. & - & (\checkmark) & (\checkmark) & (\checkmark) \\
\\
\midrule

\end{longtable}

\section{Definitions of Stakeholder Roles}
\label{appendix:roles}

\begin{longtable}
{p{0.3\linewidth}
p{0.7\linewidth}}
\hline
\textbf{Role} & \textbf{Definition} \\
\hline
\endfirsthead

\hline
\textbf{Role} & \textbf{Definition} \\
\hline
\endhead

\hline
\endfoot

\hline
\endlastfoot

\textbf{Domain Scientists (Ecologists)} &
Scientists who define the scientific objectives, assumptions, and conceptual scope of environmental models, ensuring alignment with real-world ecological processes and DT goals. \\
\hline

\textbf{Measurement Model Designers} &
Experts who design observation, measurement, and monitoring models based on scientific requirements, including variables, scales, and methodological assumptions. \\
\hline

\textbf{Data Curators} &
Individuals responsible for verifying data quality, annotating datasets, managing metadata, versioning data products, and ensuring long-term preservation and usability. \\
\hline

\textbf{Semantic Curators} &
Specialists who design and maintain conceptual and semantic models, vocabularies, and ontologies, enabling semantic alignment and interoperability across models and datasets. \\
\hline

\textbf{Data Scientists} &
Professionals who analyze, transform, and interpret data using computational and statistical methods, often integrating models and datasets within analytical workflows. \\
\hline

\textbf{Software Developers} &
Engineers who implement, adapt, and maintain computational models, interfaces, workflows, and integration components required for DT execution. \\
\hline

\textbf{Infrastructure Managers} &
Professionals responsible for operating and maintaining the computational infrastructure supporting DT systems, including hardware, platforms, networks, and execution environments. \\
\hline

\textbf{Platform Operators} &
Actors responsible for managing virtual research environments or modeling platforms, including user access, service availability, catalogues, and operational stability. \\
\hline

\textbf{DT Integrators} &
Specialists who design and implement the integration of heterogeneous models, datasets, and services into coherent DT systems, resolving compatibility issues across semantic, technical, and runtime dimensions. \\
\hline

\textbf{Asset Providers} &
Organizations or platforms that supply reusable digital assets such as datasets, models, software components, or infrastructure services for DT development. \\
\hline

\textbf{FAIRification Stewards} &
Individuals responsible for ensuring that data and models comply with FAIR principles, enhancing their findability, accessibility, interoperability, and reusability. \\
\hline

\textbf{Policymakers} &
Stakeholders who use DT outputs to support decision-making, policy design, and evaluation, often requiring transparency, provenance, and reliability guarantees. \\
\hline

\textbf{Partners and Collaborators} &
Organizations or individuals contributing complementary expertise, resources, or assets to DT development through collaborative arrangements. \\
\hline

\textbf{Users} &
Researchers or practitioners who compose workflows, run models, and interact with DT systems to conduct scientific analysis or exploratory studies. \\
\hline

\end{longtable}




\end{appendices}


\bibliography{mybibfile}


\end{document}